\documentclass[prd,aps,nofootinbib]{revtex4}
\usepackage{slashed}
\usepackage{graphicx}
\usepackage{amsmath}
\usepackage{amssymb}
\usepackage{amsfonts}
\usepackage{indentfirst}
\usepackage{color}
\newcommand{\nin}{\noindent}
\newcommand{\be}{\begin{equation}}
\newcommand{\ee}{\end{equation}}
\newcommand{\bea}{\begin{eqnarray}}
\newcommand{\eea}{\end{eqnarray}}

\newcommand{\nn}{\nonumber\\}

\newcommand{\ovl}{\overline}

\begin{document}

\preprint[\leftline{KCL-PH-TH/2019-{\bf 46}}

\title{\Large {\bf Weak-U(1)~$\times$~Strong-U(1) Effective Gauge Field Theories  and electron-monopole scattering}}

\bigskip

\author{Jean Alexandre and Nick E. Mavromatos}

\bigskip

\affiliation{Theoretical Particle Physics and Cosmology Group, Physics Department, King's College London, Strand, London WC2R 2LS, UK.}

 \begin{abstract}
\vspace{0.5cm}

We present a gauge and Lorentz invariant model for the scattering of matter off magnetic poles, which justifies the 
presence of velocity-dependent magnetic charges as an effective description of either the behaviour of monopoles in scattering with matter or their 
production from matter particles at colliders. Hence, in such an approach, perturbativity of the magnetic charge is ensured for relative low velocities of 
monopoles with respect to matter particles. The model employs a ${\rm  U(1)}_{\rm weak} \times {\rm U(1)}_{\rm strong}$ effective gauge field theory under 
which electrons and monopoles (assumed to be fermions) are appropriately charged. The non-perturbative quantum effects of the strongly coupled sector 
of the theory lead to dressed effective couplings of the monopole/dyon with the electromagnetic photon,  due to non-trivial wave-function renormalization effects.
For slowly-moving monopole/dyons, such effects lead to weak coupling, thus turning the bare non-perturbative magnetic charge, which is large due to the Dirac/Schwinger 
quantization rule, into a perturbative effective, velocity-(``$\beta$'')-dependent magnetic coupling. Our work thus offers formal support 
to previous conjectural studies, employing effective U(1)-electromagnetic gauge field theories for the description of monopole production from Standard Model matter, 
which are used in contemporary collider searches of such objects. This work necessarily pertains to composite monopoles, 
as seems to be the case of all known  monopoles so far,  that are solutions of specific particle physics models. This is a consequence of the fact that the wave-function 
renormalization of the (slowly-moving) monopole fermion turns out to violate unitarity bounds that would characterise asymptotic elementary particle states.
\end{abstract}

\maketitle

\section{Introduction and Motivation \label{sec:intro}} 

The quantum theory of structureless magnetic poles as introduced by Dirac~\cite{Dirac}, was characterised by non-local hidden degrees of freedom, the ``Dirac string'', whose 
invisibility in physical processes lead to the charge quantization condition. The presence of the string lead also to Lorentz non-invariance of the configuration and the 
associated field theory treatments, which manifested itself in various contexts, such as  the local formulation of the magnetic charges by Zwanziger~\cite{Zwanziger:1970hk}, 
which necessitated the presence of a fixed four vector in the associated effective Lagrangian, containing two gauge potentials, associated with electric and magnetic current sources, or the 
famous Weinberg's paradox~\cite{Weinberg}, according to which the leading perturbative term in the scattering amplitude between en electric charge and a magnetic pole was not Lorentz invariant. 
Schwinger's formulation of the dyon~\cite{schw}, which generalised the magnetic monopole to an object carrying both electric $q_e$ and magnetic $q_m$ 
charge, restored Lorentz invariance, but at the cost of  introducing a non-local Hamiltonian formulation of the dyon field, provided Schwinger generalisation of Dirac's 
quantization condition is valid, for the scattering of dyon configurations with charges $q_e^n$, $g_m^\ell$, $n, \ell$ positive integers:
\begin{equation}\label{schquant}
\Big(q_e^n\, g_m^\ell  - q_e^\ell \,g_m^n \Big)/4\pi = {\mathcal Z}_{n\ell} \in {\tt Z}~, 
\end{equation}
where ${\tt Z}$ denotes the set of integers.
In the absence of electric charges, this condition leads to the Dirac quantisation (but the fundamental charge unit is twice that of Dirac~\cite{Dirac}). 
It should be remarked that, upon the imposition of \eqref{schquant}, any Lorentz non-invariant effect in the effective two-gauge-potential Lagrangian of 
Zwanziger~\cite{Zwanziger:1970hk} disappears, and this is also associated with an integrability condition of the representation of the Poincar\'e Lie algebra, that stems from Poincar\'e invariance into a representation of the finite Poincar\'e group~\cite{Zwanziger:1970hk}.

Although the initial concept of the monopole as envisaged by Dirac was structureless, subsequently, `t Hooft and Polyakov~\cite{thooft}
proposed {\it composite} monopoles, which were {\it topological soliton} solutions of phenomenologically realistic 
gauge and Lorentz invariant field theories, involving spontaneous (Higgs-like) symmetry breaking. 
Unlike the Dirac monopole, such solutions were smooth, with {\it no} Dirac-string singularities, being characterised by finite energy.\footnote{The fact that such 
solutions were obtained from gauge and Lorentz invariant field theories, made Weinberg's observation on the non-Lorentz invariant nature of the (perturbative) scattering 
of electric charges off (generic) magnetic poles, a true paradox.}
The solution is localised around the origin, where the gauge group is unbroken, but very far from it the gauge group G is broken to a subgroup H, 
and the `t Hooft-Polyakov monopole reduces to the Dirac one.  `t Hooft considered the Georgi-Glashow model~\cite{gg} involving a 
Higgs triplet that spontaneously breaks the SU(2) group. The importance of this type of simply connected gauge groups is that they are characterised by a non-trivial homotopy, 
e.g. for the SU(2) gauge group $\pi_2$(SU(2))$= {\tt Z}$, with ${\tt Z}$ the set of integers. The defines how many times the spatial three-sphere, where the Higgs field lives on, 
wraps around the internal (gauge) space sphere spanned by the Higgs triplet of the Georgi-Glashow model, which leads to the magnetic charge quantization condition (\eqref{schquant} 
for the magnetic monopole, i.e. $q_e^\ell=0$). Extension of the monopole and dyon solutions to phenomenologically realistic Grand Unified Theories (GUT) with gauge group SU(5), 
with similar conclusions,  were made in \cite{dokos}. In such models, the monopoles have masses of order of the GUT scale $10^{14}-10^{16}$ GeV, which implies that 
inflation (whose scale is believed to be closed to this scale) would have inflated them away, which probably explains why such large GUT scale monopoles have not been discovered in 
cosmic searches so far~\cite{patrizii}.  In superstring theories, with appropriate grand unifying gauge groups, broken by stringy methods (Wilson lines), structures with magnetic 
charges have been first considered in \cite{wen}. Moreover, in D-brane-inspired GUT models~\cite{shafi}, the unification scale (and hence the mass of the appropriate magnetic monopoles/dyons)  
 can be lowered significantly, down to $10^4-10^6$ GeV,  relevant for future collider or cosmic-ray searches. 

Unfortunately, unlike the SU(2) or SU(5) or other GUT-like-group monopoles, the gauge group of the standard model (SM) SU(2)$\times$ U$_Y$(1), does not have this  simple structure allowing for topological charge quantization, because of the hypercharge U$_Y$(1) factor. As a result, after Higgs breaking, the quotient group SU(2)$\times$ U$_Y$(1)/U$_{\rm em}$(1) is not characterised by a non-trivial second homotopy, thus monopoles were not expected to exist in the Standard Model. 
Phenomenologically, such a conclusion would imply that  current collider (LHC) monopole searches might be futile. 

However, in \cite{cho}, it was argued that  one can look for non-trivial topology in the Higgs field structure. Indeed, in the presence of the U$_Y$(1) hypercharge group factor, the Weinberg Salam model is viewed as a gauge $CP^1$ model with the (normalised) Higgs doublet field playing the role  of the corresponding CP$^1$ field, characterised by a non-trivial homotopy $\pi_2(CP_1)={\tt Z}$ thus allowing in principle for a topological quantization \`a la `t Hooft-Polyakov. This monopole is not characterised by a Dirac string, given that the latter had been argued to be a gauge artifact, provided the U$_Y$(1) hypercharge group is endowed with a non-trivial bundle structure~\cite{cho}. Unfortunately, however, the resulting monopole or dyon solutions of \cite{cho} were characterised by infinite energy. 

Recently, though, it was observed that finite energy monopoles of the type proposed in \cite{cho} can characterise extensions of the Standard Model, with either appropriate non-minimally coupled Higgs and hypercharge sectors~\cite{emy}, or higher-derivative extensions of the hypercharge sector, for instance in a (string-inspired) Born-Infeld configuration~\cite{aruna}. Such monopole/dyon  solutions  could have masses accessible to the scales of current or future colliders. 
Other finite-energy structured monopole/dyon solutions with potentially low mass (characterised by a Dirac string, though) can be found in string-inspired models with axion-like structures~\cite{sarkar}.

In view of the above theoretical advances in the field of relatively light monopoles/dyons,  the development of effective field theory methods that could allow study of their production  at colliders or scattering off SM matter has gained renewed interest. So far, lacking a fundamental theory for the description of the interaction of magnetic poles with standard model matter, like leptons, quarks and photons, only {\it ad hoc} phenomenological effective U(1) gauge field theory models are used for collider searches of monopoles~\cite{monprod}.  

These models are essentially dual extensions of electrically charged particles of various spins interacting with photons, in which the electric charge in the coupling with the photon is replaced by an effective magnetic charge. Such effective descriptions may contain some truth in them, provided one considers the above-mentioned cases of the composite monopole solutions in (extensions of) the SM. Indeed, as discussed in \cite{drukier}, one may think of the magnetic monopole charge in such cases as a 
{\it collective coupling} to photons of (electrically) charged constituent degrees of freedom, such as charged $W$-bosons and Higgs fields, which the monopole is composed of. Modeling these constituent fields as quantum harmonic oscillators, the authors of \cite{drukier} argued that the 
monopole might be viewed as a {\it coherent superposition} of 
$\sim \frac{1}{\alpha}$ such quantum states, with the result that the collective coupling to photons is $\frac{1}{\alpha} \, e$, consistent with the lowest non-trivial sector of the charge 
quantization condition \eqref{schquant}. 

In this work we shall attempt to construct (strongly coupled) effective gauge field theories for the above type of composite monopoles, by extending non trivially the ideas of 
Zwanziger~\cite{Zwanziger:1970hk} that were developed for structureless monopoles. 
When we consider our effective theory sufficiently far away from the monopole centre, the `t-Hooft-Polyakov-type monopoles mentioned above resemble the structureless Dirac ones, to a good approximation. Nonetheless, as we shall explain below, 
the role of compositeness will turn out to be crucial for us, since, as we shall see, when we consider the  quantisation of our non-perturbative effective field theory, the resulting wave-function renormalization  for the monopole field will not respect the appropriate unitarity bounds for an elementary field, thus making our effective description suitable only for composite fields~\cite{psbook}. We shall be quite generic though in our considerations, and we shall  not specify the type of the composite monopole/dyon. For us, it might be 
one of the known types, mentioned above, or an as yet unknown solution of some beyond the SM theory with or without Dirac string singularities.

Let us commence our study by first reviewing Zwanziger's approach \cite{Zwanziger:1970hk}, which we shall make partial use of in this work. 
The approach employs two related gauge fields, which allows for the construction of a local Lagrangian for the description of dyons, albeit non Lorentz invariant. 
If one considers electric and magnetic currents, $J_e^\mu$ and $J_m^\mu$ respectively,  
 then, as shown in \cite{Zwanziger:1970hk}, the corresponding Maxwell's equations read
 \begin{equation}\label{maxwell}
 \partial_\mu \, F^{\mu\nu} = J_e^\nu, \quad \partial_\mu \, ^\star F^{\mu\nu} = J_m^\nu, 
 \end{equation}
where $F^{\mu\nu}$ denotes the electromagnetic field-strength tensor, with $^\star F^{\mu\nu} \equiv \frac{1}{2}\epsilon^{\mu\nu}_{\,\,\,\,\rho\sigma}\, F^{\rho\sigma}$  the dual tensor, and 
$\epsilon^{\mu\nu\rho\sigma}$  the totally antisymmetric Levi-Civita symbol, with $\epsilon^{0123}=+1$, etc. We work throughout this work in a flat Minkowski 
space-time with metric $\eta^{\mu\nu}=(1, -1, -1, -1)$. For future purposes, we note the {\it axial}-vector (pseudovector) nature of the magnetic current in \eqref{maxwell}.

The general solution of the two Maxwell's equations, can be expressed in terms of two (related, as we shall see)
potentials $\mathcal A_\mu$ and $\mathcal B_\mu$
and a fixed four vector $\eta^\mu$, as follows~\cite{Zwanziger:1970hk}:
\begin{equation}\label{gensol} {\rm First~eq.~with~electric~current:} \, F = - ^\star (\partial \wedge \mathcal B) + (\eta \cdot \partial)^{-1} \, 
(\eta \wedge J_e), \quad ^\star F= \partial \wedge \mathcal B + (\eta \cdot \partial)^{-1} \, ^\star (\eta \wedge J_e), 
\end{equation}
\begin{equation}\label{gensol2}
{\rm Second~eq.~with~magnetic~current:} \, ^\star F = ^\star (\partial \wedge \mathcal A) + (\eta \cdot \partial)^{-1} \, (\eta \wedge J_m), 
\quad F= \partial \wedge \mathcal A - (\eta \cdot \partial)^{-1} \, ^\star (\eta \wedge J_m),
\end{equation}
where we used a form notation for brevity, with $\wedge$ ($\cdot$) denoting exterior (interior-dot) product as usual, such that for four vectors 
$(a \wedge b)^{\mu\nu} \equiv a^\mu b^\nu - a^\nu b^\mu$, $a \cdot b \equiv a^\mu b_\mu$.   With these conventions we have, for any antisymmetric second-rank tensor 
$\mathcal F_{\mu\nu}=\mathcal F_{\nu\mu} $: $^\star \,^\star \mathcal F_{\mu\nu} = - \mathcal F _{\mu\nu}$. 

As discussed in \cite{Zwanziger:1970hk}, one can eliminate the currents from \eqref{gensol} and \eqref{gensol2}, 
and express these equations solely in terms of the potentials $\mathcal A_\mu$ and $\mathcal B_\mu$. The following representation of the kernel $(\eta \cdot \partial)^{-1}(x)$ (satisfying 
$\eta \cdot \partial \, (\eta \cdot \partial)^{-1}(x) = \delta^{(4)}(x)$) is used:
\begin{equation}\label{kernel}
(\eta \cdot \partial)^{-1}(x) = c_1 \int_0^\infty \delta^{(4)}(x - \eta \, s) \, ds - (1 - c_1) \,  \int_0^\infty \delta^{(4)}(x +\eta \, s) \, ds,
\end{equation}
with $c_1$ a real constant,  appropriately defined in order to obtain the correct form of the Lorentz force  in the classical relativistic particle limit of the dyon field~\cite{Zwanziger:1970hk}. 
The form \eqref{kernel} implies that, in the point-particle case, the support of $(\eta \cdot \partial)^{-1}(x_i - x_f)$ is reduced to $x_i^\mu (\tau_i) -  x_i^\mu (\tau_f) = \eta^\mu s$, 
for $-\infty < \tau_i, \tau_f, s < +\infty$, with $\tau$ the proper time.

The gauge potentials $\mathcal A_\mu$ and $\mathcal B_\mu$ depend on $\eta^\mu$ and on the gauge choice. For convenience, in the approach of ref.~\cite{Zwanziger:1970hk}, 
the fixed four vector $\eta^\mu$ was chosen to be space-like $\eta^\mu \eta_\mu < 0$. The gauge potentials are not independent, as they are associated with a single field strength $F$, 
since the dual $^\star F$ is expressed in terms of $F$. 
Indeed, from \eqref{gensol}, by equating the expressions for the field strength $F$ between the first equation in \eqref{gensol} and the second equation in \eqref{gensol2}, yields:
\begin{equation}\label{relAB}
\partial \wedge \mathcal A + ^\star (\partial \wedge \mathcal B) = (\eta \cdot \partial)^{-1} \Big[ \eta \wedge J_e + ^\star (\eta \wedge J_m)\Big],
\end{equation}
 so that only two photon degrees of freedom propagate on-shell in this local theory, despite the fact that two potentials are needed to ensure a local formulation of the dyon. 
 The constraint \eqref{relAB} would imply that $\mathcal B$ is an {\it axial vector}, since this is the case with the magnetic current as well, and in this way one obtains consistent 
 transformations under the (improper) Lorentz group, involving reflexions. This will be essential for our purposes in this work, when we formulate in the next section the effective 
 quantum gauge field theory for the monopole. 
 
 We also remark at this point that, upon using the following identity for an antisymmetric second rank tensor ${\mathcal F}_{\mu\nu}=-{\mathcal F}_{\nu\mu}$:
\begin{equation}\label{identity}
{\rm tr}({\mathcal F} \cdot {\mathcal F}) \equiv {\mathcal F}_{\mu\nu} {\mathcal F}^{\nu\mu} =\frac{2}{\eta^2} \Big[ -(\eta \cdot {\mathcal F})^2 +(\eta \cdot ^\star {\mathcal F})^2\Big],
\end{equation}
with the notation $\eta^2 = \eta_\mu \eta^\mu$, $(\eta \cdot {\mathcal F})^\nu = \eta_\mu {\mathcal F}^{\mu\nu}$, etc., it is possible 
express the electromagnetic tensor $F$ and its dual $^\star F$, \eqref{gensol}, \eqref{gensol2}, in terms of the potentials $\mathcal A_\mu$ and $\mathcal B_\mu$ alone~\cite{Zwanziger:1970hk}:
\be\label{ffd}
F = \frac{1}{\eta^2} \Big( \eta \wedge (\eta \cdot [\partial \wedge {\mathcal A}]) - ^\star \{\eta \wedge (\eta \cdot [\partial \wedge {\mathcal B}])\}\Big), \quad 
^\star F  = \frac{1}{\eta^2} \Big( \, ^\star\{\eta \wedge (\eta \cdot [\partial \wedge {\mathcal A}])\} + \{\eta \wedge (\eta \cdot [\partial \wedge {\mathcal B}])\}\Big),
\ee
in differential form notation.

Taking into account the identity \eqref{identity}, the Lagrangian yielding the field equations \eqref{maxwell}, in agreement with the solutions \eqref{gensol}, \eqref{gensol2}, 
reads~\cite{Zwanziger:1970hk}:
\begin{eqnarray}\label{lag}
{\mathcal L} &=& \frac{1}{8}{\rm tr}\Big[(\partial \wedge \mathcal A) \cdot (\partial \wedge \mathcal A)\Big] +  
\frac{1}{8}{\rm tr}\Big[(\partial \wedge \mathcal B) \cdot (\partial \wedge \mathcal B)\Big] - J_e \cdot \mathcal A - J_m \cdot \mathcal B \nonumber \\
&-& \frac{1}{4\eta^2}\, \Big(\eta \cdot \Big[(\partial \wedge \mathcal A)  + ^\star (\partial \wedge \mathcal B)\Big]\Big)^2  - 
\frac{1}{4\eta^2}\, \Big(\eta \cdot \Big[(\partial \wedge \mathcal B)  - ^\star (\partial \wedge \mathcal A)\Big]\Big)^2~.
\end{eqnarray}
The expression \eqref{lag} has the advantage of separating, formally at least, the Lorentz-violating effects of the fixed four-vector $\eta$ from the conventional Lorentz-invariant form 
of the Quantum Electrodynamics (QED) Lagrangian in terms of the photon field. 
Indeed, the first line of \eqref{lag} would correspond  to the conventional (Lorentz invariant, isotropic) QED terms, independent of the four-vector $\eta^\mu$, while the  
second line would correspond to the Lorentz-violating effects of the Dirac string
in the dyon/monopole case, under the assumption that the Dirac string direction is aligned with that of the fixed space-like four-vector $\eta^\mu$. 
The presence of the monopole, and its topologically non-trivial nature (the solutions belong to a field-theory 
sectors of monopole number $n =1,2, \dots$) is reflected precisely on the impossibility to deform continuously the vector 
so as $\eta^\mu \to 0^\mu \equiv (0,0,0,0)^T$, with $T$ denoting matrix transposition. However, by using the form \eqref{lag}, such a limit can be {\it formally} taken. 
Indeed, eq.~\eqref{relAB} implies in such a case 
\be\label{solconstr}
\partial \wedge \mathcal A  + ^\star (\partial \wedge \mathcal B)\stackrel{\eta^\mu \to 0^\mu}{=} 0~, 
\ee
since, in view of \eqref{kernel}, we can formally set the $\eta$-dependent right-hand-side to zero (given that it approaches $0$ faster than $\eta^2 \to 0$, 
since the operator $(\eta \cdot \partial)^{-1}(x_i - x_f)$ has zero support in the limit $\eta^\mu \to 0^\mu$). The Lagrangian (\ref{lag}) then does not contain any 
Lorentz-symmetry violating term, and involves two gauge fields related by the constraint (\ref{solconstr}). Our starting point is a similar model, as we explain in the next section.

 In \cite{terning}, it was argued, by means of topological considerations, 
that any Lorentz-violating effects of a magnetic pole (due to the Dirac string), can be resummed  in a non-perturbative way in such a way that 
the scattering amplitude of an electric charge off a magnetic charge contains all such Lorentz-violating effects in a phase, which thus drops out of physical quantities such as cross sections. 
Moreover, this phase turns out to be a multiple of $2\pi$, provided the quantization condition \eqref{schquant}, and thus in such a case the amplitude is Lorentz invariant. 
Such a conclusion was reached by means of studying a toy model employing perturbative magnetic charges in a dark sector, using the two potential formalism \eqref{lag} appropriately 
in both the visible and dark sectors, and assuming a perturbative small mixing of ordinary photons with dark photons, that leads to perturbative couplings of magnetic charges to 
ordinary photons in the visible sector. This allows for a perturbative resummation of the ``dark'' monopole effects, making use of appropriate soft emissions of both gauge fields 
$A^\mu$ and $B^\mu$ in the pertinent Feynman diagrams, describing the scattering of electrons off monopoles, which in turn leads to the aforementioned decoupling of the 
Lorentz-violating-(Dirac-string-like) effects of the vector $\eta^\mu$ from the relevant cross sections.

Before proceeding further, we would like to mention another peculiarity of the dyon, discussed in \cite{milton}, which will turn out to be of pivotal importance for our purposes here. 
When considering dyon-dyon 
scattering within a non-relativistic quantum mechanical framework, as appropriate for small relative velocities of the scattered dyons, 
Schwinger {\it et al.} arrived at the following differential cross section 
for the particular case of an electric charge $e$ scattered off a magnetic one $g$ (in units with the speed of light in vacuo is $c=1$):
\be\label{diffcs}
\frac{d\sigma}{d\Omega} \simeq \left(\frac{eg}{2\mu v}\right)^2\frac{1}{(\theta/2)^4}
\ee
where $\mu$ is the reduced mass, $\theta$ is the (small) scattering angle of the non relativistic scattering, and $v = |\vec v|$ is the magnitude of relative velocity $\vec v$ of the 
magnetic charge . 

If the magnetic pole is a dyon, carrying also an electric charge $e_d$, then (\ref{diffcs}) is extended {\it classically} to~\cite{milton}
\be\label{ddcs}
\frac{d\sigma}{d\Omega} \simeq \left(\frac{1}{2\mu v}\right)^2 \, \Big[ (e\, g)^2 + \frac{(e\, e_d)^2}{v^2} \Big]\, \frac{1}{(\theta/2)^4}~.
\ee
However, when consider {\it quantum scattering}, of an electron off a dyon, the small-angle formula for the differential cross section contains~\cite{milton} only 
the first term inside the brackets of \eqref{ddcs}
and thus {\it coincides} with \eqref{diffcs}. This will turn out to be important for our work, when we define the magnetic charge in section \ref{emscat} using scattering arguments.

One observes that the cross section \eqref{diffcs} of electron-magnetic-monopole scattering can be obtained from the Rutherford differential cross section 
\be\label{Rutherford1}
\left.\frac{d\sigma}{d\Omega}\right|_{Ruth}=\left(\frac{e^2}{2\mu v^2}\right)^2\frac{1}{(\theta/2)^4}
\ee
upon the replacement $e^2\to e\, g_{eff}$, where the effective monopole charge is (we reinstate the units of $c$ in this formula for notational clarity, 
so that the reader connects easily to the existing literature~\cite{milton})
\be\label{geff1}
g_{eff}\equiv g \frac{v}{c} \equiv g \, \beta.
\ee
Upon invoking {\it electric-magnetic duality}, one therefore might expect that $g_{eff}$ defines an effective ``velocity-dependent'' magnetic charge that describes the 
behaviour of a magnetic monopole in matter (or equivalently its production from the collision of matter Standard model particles, 
such as quarks or charged leptons, at colliders), The important thing of having a magnetic coupling \eqref{geff1} is its perturbative nature for small $v$, which can be used in 
monopole searches at colliders to place monopole mass bounds~\cite{moedal}. In fact, in the case of monopole-antimonopole pair production mechanisms at colliders, one 
may use such effective magnetic charges but in a Lorentz-invariant manner, using the centre-of-mass velocity~\cite{monprod}
\begin{equation}\label{defbeta}
\beta = \sqrt{1 - \frac{4M^2}{s}},
\end{equation}
where $M$ is the monopole mass, and $s=(p_1+p_2)^2$ is the Mandelstam variable, with $p_i$, $i=1,2$, the momenta of the colliding particles (quarks of photons, 
in Drell-Yan or photon fusion processes for monopole production, respectively).

It is the purpose of this work to justify the use of an effective magnetic charge \eqref{geff1} in scattering processes involving magnetic poles by constructing 
a Lorentz and gauge invariant effective field theory of monopoles within a toy model, which we shall describe below. Our model is motivated by the model of 
Zwanziger~\cite{Zwanziger:1970hk} \eqref{lag}, combined with the findings/arguments of \cite{terning} that any Lorentz-Violating effects of the monopole will not be present in cross 
sections (or scattering amplitudes, if the quantization condition \eqref{schquant} is valid).  Our results are based on a non-perturbative dressing of 
the coupling of the monopole to the real photon. At present, it is not technically possible to work  directly with the 
Lagrangian \eqref{lag}, due to complications arising from the presence of the four vector $\eta^\mu$, and the associated constraint \eqref{relAB}. 
However, as our main purpose here is to demonstrate the emergence of non-perturbative dressing of the magnetic coupling, 
we deviate from the letter of the Zwanziger approach~\cite{Zwanziger:1970hk}, keeping as much as possible of its spirit. 

To this end, we consider a model Lagrangian, involving two independent U(1) 
gauge potentials $A_\mu, B_\mu$, with $A_\mu$ denoting the ordinary photon of the weakly coupled quantum electrodynamics (denoted by U(1)$_{\rm weak}$). 
The gauge field $B_\mu$, referred to as a `dual photon', belongs to the Lie algebra of a strongly coupled U(1)$_{\rm strong}$, which is independent of that of electromagnetism. 
We also ignore any Lorentz violating effects, anticipating
the results of \cite{terning}, based on soft gauge-field resummation, which we discuss briefly at the end of our article. 
The ordinary standard matter is represented as a spin-1/2 fermion, for simplicity, carrying electric charge only. For concreteness, in this work we assume the matter fermion 
to be an electron of charge $e$, but extension to any other electrically charged fermion in the Standard Model sector is straightforward. 
The dyon, on the other hand, is represented as a dual 
of the ordinary electron, under the aforementioned electric-magnetic duality, which lead to the symmetry of Maxwell's equations if magnetic poles are present. 
Hence it is a fermion field itself (spin-1/2), which however carries {\it both} electric and magnetic charge, and thus couples to both $A_\mu$ and $B_\mu$ fields. 
Our approach consists in studying coupled Schwinger-Dyson equations, so as to study the effective vertex of the coupling of the dyon to the real photon which, as we shall demonstrate, 
can be identified with an effective magnetic charge of the form \eqref{geff1}, consistently with electric-magnetic duality.

The structure of the article is the following: in the next section \ref{sec:model}, we present the U(1)$_{\rm weak} \times $ U(1)$_{\rm strong}$ - model, 
discuss its properties and present the relevant set of coupled Schwinger-Dyson (SD) equations. In section \ref{nonperturb} we solve these equations in 
the non-perturbative regime for the coupling of the dual photon, but in the weak-coupling limit of the ordinary quantum electrodynamics (QED), and obtain expressions 
for the dressed dual photon  propagator and the photon-monopole vertices. Of particular interest to our study is the ordinary-photon-monopole dressed vertex, which 
turns out to be proportional to the monopole-field wave-function renormalization. The latter is computed in a self-consistent way in the strong coupling limit of the dual U(1) 
gauge theory, under the assumption that it is approximately  momentum independent. The gauge independence of the associated physical observables is also discussed. 
In the following section \ref{emscat}, we discuss electron (matter)-monopole scattering, compute the relevant cross section and derive the effective velocity-dependent magnetic charge 
\eqref{geff1} as a dressed coupling of the monopole-real-photon vertex derived previously. An important role in this is played by the appropriate physical 
interpretation of the dimensional-transmutation scale than enters the expression for the wave-function renormalization within our dimensional-regularisation treatment.
For slowly-moving monopoles, the resulting wave-function renormalization will turn out to violate unitarity bounds for elementary particle states. 
Thus, our effective field theory can {\it only} describe {\it composite} monopoles. 
Finally, section \ref{concl} contains our conclusions and outlook. Technical aspects of our analysis are given in several Appendices.

\section{The Model \label{sec:model}}

We consider a gauge field theory with two independent sectors: 
{\it(i)} the Standard Model, with electric charges only and a corresponding U(1)$_{\rm weak}$ gauge field $A_\mu$; 
{\it(ii)} the monopole/dyon sector, ``electrically'' charged under $A_\mu$ (with a bare charge $e_A$) and ``magnetically'' coupled to an independent U(1)$_{\rm strong}$ axial 
gauge field $B_\mu$ (with the corresponding bare coupling $e_B$). 
Before proceeding, we should make some important comments regarding the characterisation of the monopole/dyon as `electrically charged' under U(1)$_{\rm weak}$.
From a naive point of view it seems that we are dealing here with a dyon in the sense of Schwinger~\cite{schw} or Zwanzinger~\cite{Zwanziger:1970hk}, whose bare electric charge is $e_A$. 
However, as we shall discuss in this work, the non-perturbative dressing of the corresponding monopole-electromagnetic-photon vertex will lead to a non-conventional dressed `electric charge'
proportional to the (square of the) `magnetic charge'  $e_B$ ({\it cf.}  \eqref{magneticcharge}, \eqref{Z2} in section \ref{emscat}, below), and thus vanishing when $e_B \to 0$. 
In this sense, this dressed coupling is the induced {\it magnetic charge coupling} of the monopole/dyon to the real photons, assumed {\it ad hoc} in the phenomenological searches 
of magnetic monopoles using effective field theories~\cite{monprod}.  Our approach provides a microscopic derivation of this effect.

For our purposes it is enough to consider a simplified matter sector, consisting of an electron $\psi$ of (bare) mass $m$, coupled to the Standard Model photon $A_\mu$ with a 
perturbative electric charge $e$. 
We also consider a spin-$\frac{1}{2}$ monopole $\chi$, of (bare) mass $M$, coupled electrically to $A_\mu$ and magnetically to 
the axial dual-photon $B_\mu$. Unlike the Zwanziger approach~\cite{Zwanziger:1970hk}, here $A_\mu$ and $B_\mu$ are independent U(1) gauge degrees of freedom, 
which suffices for our main purpose, which is a demonstration of the emergence of an effective velocity-dependent coupling of the monopole to the ordinary photon, 
after non-perturbative quantum dressing. As already mentioned, the restriction to a fermion monopole makes direct contact with the work of \cite{Zwanziger:1970hk} 
and is motivated by the electric-magnetic duality, under which the dual of an electron 
is such a fermion monopole, although it must be said that consistent formulations of field theories of monopoles with spin one or zero might also exist.

\subsection{Lagrangian}

The Lagrangian of our model is given by
\be\label{Lagrangian}
L=-\frac{1}{4}F_A^{\mu\nu}F^A_{\mu\nu}-\frac{1}{4} F_B^{\mu\nu}~ F^B_{\mu\nu}+\ovl\psi(i\gamma^\mu D^A_\mu-m)\psi+\ovl\chi(i\gamma^\mu D^{A+B}_\mu-M)\chi~,
\ee
where $A_\mu$ is a vector and $B_\mu$ is an {\it axial} vector (pseudovector). The corresponding field strength tensors are given by
\be
F_A^{\mu\nu}=\partial^\mu A^\nu-\partial^\nu A^\mu~~~,~~~~F_B^{\mu\nu}= \partial^\mu B^\nu -\partial^\nu B^\mu~,
\ee
and the covariant derivatives are:
\be
D^A_\mu=\partial_\mu-ie A_\mu~~~~\mbox{and}~~~~D^{A+B}_\mu=\partial_\mu-ie_A A_\mu-ie_B B_\mu~,
\ee
where $e$ is the coupling the electron to the gauge field $A_\mu$ and $e_A,e_B$ are the couplings of the monopole to $A_\mu$ and $B_\mu$ respectively.
These covariant derivatives ensure that the Lagrangian is invariant under the gauge transformation
\bea\label{gauge}
A_\mu&\to& A_\mu+\partial_\mu\theta_A\\
B_\mu&\to& B_\mu+\partial_\mu\theta_B\nn
\psi&\to& \exp(ie\theta_A)~\psi\nn
\chi&\to& \exp(ie_A\theta_A+ie_B\theta_B)~\chi~.\nonumber
\eea
Few remarks are in order at this point:
\begin{itemize}
 \item The axial nature of the gauge field $B_\mu$ is required for consistency of the field equations ( see eq.~(\ref{newmax}) below), under improper Lorentz transformations, 
 including spatial reflexions and reversal in time. As a consequence, the Lagrangian (\ref{Lagrangian}) breaks parity, P, and time reversal symmetry, T, but preserves CPT, 
 where C denotes charge conjugation. It should be noted that such an explicit parity and time reversal symmetry breaking, but CPT conservation, is a generic feature of theories 
 with magnetic charges~\cite{cptmag}.
 \item The (bare) electric charges $e$ and $e_A$ are both assumed to be perturbative: $e\ll1$ and $e_A\ll1$; as already mentioned, we assume here, for concreteness, 
 that the  electrically charged matter fermion $\psi$ is an electron of charge $e$. Our analysis can be of course extended trivially to incorporate any other charged matter, 
 in which case the coupling $e$ will be replaced by the corresponding  electric charge $q_e$. 
 \item The coupling $e_B$ of the spin-$\frac{1}{2}$ monopole to the dual photon should not be identified immediately with the magnetic charge of the monopole/dyon. 
 The latter will be defined appropriately later on, in section \ref{emscat}, via studying the scattering process of dyons with matter fermions in our effective theory, 
 and identifying the relevant cross sections in the non-relativistic limit with the corresponding ones in the quantum-mechanical approach of \cite{milton}. It is this charge that 
 satisfies the quantization condition \eqref{schquant}. 
As we shall discuss in this work, this quantization condition implies strong coupling for the dual photons $e_B \gg e_A$, and it is the dressing of the electric vertex of the monopole 
with the real photon by {\it non-perturbative} quantum corrections of the strongly-coupled dual photon that leads in general to non-perturbative electromagnetic couplings of the 
monopole/dyons to the real photon. This coupling will be identified with the ``magnetic charge coupling $g$'' of the monopole to photons, appearing in phenomenological effective 
field theories used in monopole/matter scattering or production of monopoles at colliders~\cite{monprod}. As we shall discuss in section \eqref{emscat},  this magnetic charge $g$, 
coincides with the one defined by Dirac~\cite{Dirac} in  the expression for the (singular) magnetic field of the monopole. In our context, it will turn out to be a product of $e_A$ 
with the monopole/dyon wave-function renormalization $Z$, $g=Z e_A$, where the factor $Z$ is due to  the (strongly-coupled, non-perturbative) quantum effects associated with the dual 
photon of  the gauge group factor $U_{\rm strong}(1)$ of our $U(1)_{\rm weak} \otimes U_{\rm strong}(1)$ effective field theory. This wave-function renormalization factor is itself a 
non-trivial function of {\it both} couplings $e_B$ and $e_A$, and under the requirement of the validity of the charge  quantization \eqref{schquant}, will turn out to be proportional 
to $e_B$ ({\it cf.} \eqref{magneticcharge},\eqref{Z2}), thus vanishing in the absence of a magnetic coupling $e_B$.
Moreover, and most importantly, as our effective field theory approach will also indicate, such magnetic charge dressed couplings can become perturbative for slowly moving monopole/dyons, 
in agreement with the electric-magnetic-duality-inspired conjecture of \cite{milton} on the emergence of an effective magnetic charge \eqref{geff1} in scattering processes of monopole/dyons 
with matter. 
We stress once again that for us, the presence of a bare electric coupling $e_A$ of the monopole associated with the weak $U(1)_{\rm weak}$ electromagnetic gauge group is essential. 

\end{itemize}

After the above necessary remarks, we are now in position to commence our study. 
Minimising the action with respect to $A_\mu$ and $B_\mu$ respectively leads to:
\be\label{ABeqs}
\partial_\mu F_A^{\mu\nu}=ej_\psi^\nu+ e_A j_\chi^\nu~~~~\mbox{and}~~~~\partial_\mu F_B^{\mu\nu}=e_B j_\chi^\nu~,
\ee
where the currents are defined here without the charges
\be
j_\psi^\nu=\ovl\psi\gamma^\nu\psi~~~~\mbox{and}~~~~j_\chi^\nu=\ovl\chi\gamma^\nu\chi~.
\ee
Contracting  eqs.(\ref{ABeqs}) with the divergence $\partial_\nu$, one can see that both vector currents are conserved individually
\be
\partial_\nu j_\psi^\nu=\partial_\nu j_\chi^\nu=0~,
\ee
which is expected, since the gauge functions $\theta_A$ and $\theta_B$ are independent.
To make the connection with Zwanziger's approach~\cite{Zwanziger:1970hk} and the (electromagnetic) field strength tensor $F$, which appears in the field equations (\ref{maxwell}), 
we require that the gauge fields $A_\mu$ and $B_\mu$ satisfy the {\it on-shell} constraints
\be\label{newmax}
F_A ^{\mu\nu} = F^{\mu\nu} ~, \quad F_B ^{\mu\nu} = ~^\star F^{\mu\nu} ~.
\ee
This leads to a condition identical to eq.~(\ref{solconstr}), since $^\star(^\star F)=-F$. The fields $A_\mu$ and $B_\mu$ in our model will be treated as independent variables in the
quantum theory, though, and the constraints (\ref{newmax}) will then characterise {\it external} photon lines with the relevant fields put on-shell. This will be implied in what follows.

Our main point in doing this is to demonstrate the existence of effective non-perturbatively-dressed  couplings of the photons $A^\mu$ and $B^\mu$ with the monopole, 
which depend on a wave-function renormalization factor ({\it cf.} \eqref{vertices} below) that vanishes for vanishing monopole velocities ({\it cf.} \eqref{Zkk0} and \eqref{cor} below), 
thus becoming perturbative for slowly-moving monopoles. This would allow for the soft-photon resummation arguments of \cite{terning} to go through, 
implying that  any $\eta^\mu$-dependent term in \eqref{lag} can be ignored, as it will contribute only to the phase of the electron-monopole scattering amplitude; 
upon the quantization condition \eqref{schquant}, such $\eta^\mu$-dependent terms in the phase would {\it vanish} , thus leaving the amplitude itself Lorentz and gauge invariant. 

With the above in mind we now proceed to study non perturbatively the model described by \eqref{Lagrangian}. 

\subsection{Properties of the quantum theory} 
 
The notations used here are defined in Appendix \ref{appA}, where the properties of the one-particle-irreducible graph generating functional $\Gamma$ are given.

\subsubsection{Current conservation and Furry's theorem}
\label{Furry}

Because of the two independent conserved currents, there is no mixing between $\psi$ and $\chi$, and the following inverse propagators vanish
\be\label{nomixpsichi}
\left.\frac{\delta^2\Gamma}{\delta\ovl\psi\delta\chi}\right|_0=0=\left.\frac{\delta^2\Gamma}{\delta\ovl\chi\delta\psi}\right|_0~,
\ee

Also, Furry's theorem~\cite{psbook}, based on charge conjugation of magnetic currents, states that any graph involving a monopole loop vanishes if
it has an odd number of dual photon insertions. This result is also valid for an axial-vector $B_\mu$, since it relies on the charge-conjugation properties of the current only. 
As a consequence, the gauge field propagator is diagonal in gauge fields space and  
\be\label{nomixAB}
\left.\frac{\delta^2\Gamma}{\delta A_\mu\delta B_\nu}\right|_0=0~,
\ee
since any corresponding graph necessarily contains an odd number of monopole/dual-photon vertices. For the same reason, the dressed  electron/dual-photon vertex also vanishes,
$\Lambda^{B\psi}_\mu=0$, which implies that the electron does not couple to a single dual photon. The electron can couple to two dual photons though, through a two-loop process which involves
a monopole loop, and it is therefore possible to detect the electromagnetic field generated by $B_\mu$ via the electron.

\subsubsection{Ward identities} 
\label{Ward}

It is shown in Appendix \ref{appB} that the usual ward identities hold for fermions: in the limit of vanishing momentum for the gauge fields, one obtains
\bea\label{W1}
\frac{1}{e}\Lambda_\mu^{A\psi}(p,0)&=&\frac{\partial G_\psi^{-1}}{\partial p^\mu}\\
\frac{1}{e_B}\Lambda_\mu^{B\chi}(p,0)&=&\frac{\partial G_\chi^{-1}}{\partial p^\mu}~,\nonumber
\eea
We also find 
\be
q^\mu\Lambda^{B\psi}_\mu(p,q)=0~,
\ee
with $q$ denoting the gauge-field momentum, which is consistent with the fact that the electron does not couple to the dual photon. 
An additional property one finds is that ( {\it cf.}  \eqref{additionalB})
\be\label{additional}
e_Bq^\mu \Lambda^{A\chi}_\mu(p,q)=e_A q^\mu\Lambda^{B\chi}_\mu(p,q)~,
\ee
showing that the interactions between the monopole and the gauge fields are dressed with the same quantum corrections.
The condition \eqref{additional} also implies that the U(1)$_{\rm strong}$ non-perturbative quantum corrections due to the dual photons induce a non-perturbative coupling 
of the monopole/dyon to the electromagnetic photon, which is an important feature for the matter-monopole/dyon scattering or production of monopole/dyons in our framework, 
as we shall discuss later in section \ref{emscat}.

\subsubsection{Schwinger-Dyson equations}
\label{SD}

Following the usual derivation of Schwinger-Dyson (SD) equations and taking into account the property (\ref{nomixpsichi}), we show in Appendix \ref{appC} that the fermion self energies satisfy 
\bea
G_\psi^{-1}-S_\psi^{-1}&=&ie\gamma^\mu G_\psi\Lambda_{A\psi}^\nu\Delta_{\mu\nu}^A\\
G_\chi^{-1}-S_\chi^{-1}&=&ie_A\gamma^\mu G_\chi\Lambda_{A\chi}^\nu\Delta_{\mu\nu}^A+ie_B\gamma^\mu G_\chi\Lambda_{B\chi}^\nu\Delta_{\mu\nu}^B~,\nonumber
\eea
where integration over repeated space-time coordinates is understood.
Because of the property (\ref{nomixAB}), the gauge propagator is diagonal, and the gauge-field self-energies satisfy
\bea
\left(\Delta^A_{\mu\nu}\right)^{-1}-D_{\mu\nu}^{-1}&=&i~\mbox{tr}\left\{e~\gamma_\mu G_\psi\Lambda^{A\psi}_\nu G_\psi+e_A~\gamma_\mu G_\chi\Lambda^{A\chi}_\nu G_\chi\right\}\\
\left(\Delta^B_{\mu\nu}\right)^{-1}-D_{\mu\nu}^{-1}&=&ie_B~\mbox{tr}\left\{\gamma_\mu G_\chi\Lambda^{B\chi}_\nu G_\chi\right\}~.\nonumber
\eea
In the next section we apply these non-perturbative equations to study a specific limit, where the monopole does not propagate.

\section{Solution of Schwinger-Dyson Equations in the Non-perturbative regime}
\label{nonperturb}

We consider a regime which is perturbative in the electric charge $e_A<<1$, but where the coupling $e_B$ is large, $e_B>>1$. Thus, quantum corrections to the electronic sector
are neglected, and the SD equations are used to calculate quantum corrections in the monopole sector only.
In what follows, we neglect the momentum dependence of quantum corrections for the calculation of loop integrals. Because of regularisation though, these corrections acquire
a scale dependence, which will be interpreted as an external momentum dependence, in graphs describing the relevant scattering process we are interested in.

\subsection{Quantum corrections}

We derive here the monopole self-energy, as well as the polarisation tensor for the dual photon. 
In a perturbative approach, these two quantities would be determined independently but, in the non-perturbative framework of SD equations, as appropriate for strong $e_B$ couplings, 
the corresponding quantum corrections are coupled and depend on one another.

The bare fermion propagators are 
\be
S_\psi=i\frac{\slashed p+m}{p^2-m^2}~~~~,~~~~S_\chi=i\frac{\slashed p+M}{p^2-M^2}~,
\ee
and we consider the approximation where the dressed fermion propagators are
\be
G_\psi\simeq S_\psi~~~~,~~~~
G_\chi=i\frac{Z\slashed p+\tilde M}{Z^2p^2-\tilde M^2}~,
\ee
where $Z$ is the monopole wave-function renormalisation and $\tilde M$ is the dressed monopole mass. If we assume that $Z$ and 
the dressed vertices $\Lambda^{B\chi}_\mu$ and $\Lambda^{A\chi}_\mu$ are momentum-independent, the Ward identities given in 
section (\ref{Ward}) imply 
\be\label{vertices}
\Lambda_\mu^{B\chi}=e_B Z\gamma_\mu~~~~\mbox{and}~~~~\Lambda_\mu^{A\chi}=e_A Z\gamma_\mu~.
\ee
In a generic covariant gauge parametrised by $\lambda>0$, the gauge propagators assume the form
\bea\label{photonprops}
\Delta^A_{\mu\nu}&\simeq&D_{\mu\nu}=\frac{-i}{q^2}\left(\eta_{\mu\nu}+\frac{1-\lambda}{\lambda}\frac{q_\mu q_\nu}{q^2}\right) \nonumber \\
\Delta^B_{\mu\nu}&=&\frac{-i}{(1+\omega)q^2}\left(\eta_{\mu\nu}+\frac{1+\omega-\lambda}{\lambda}\frac{q_\mu q_\nu}{q^2}\right)~,
\eea
where $\omega$ is the quantum correction responsible for the dual photon transverse polarisation tensor, defined by
\be
\Delta^{B~-1}_{\mu\nu}-D^{-1}_{\mu\nu}=i\omega\left(q^2\eta_{\mu\nu}-q_\mu q_\nu\right)~,
\ee
and is also assumed to be momentum independent.

We show in Appendix \ref{appD} that the SD equation for the monopole self-energy, given in section \ref{SD}, leads to
\be\label{M=}
\frac{\delta M}{\tilde M}=\frac{1}{8\pi^2\lambda Z}\left(e_A^2(1+3\lambda)+e_B^2\frac{1+3\lambda+\omega}{1+\omega}\right)~
\frac{1}{\epsilon}\left(\frac{Zk}{\tilde M}\right)^\epsilon~+~\mbox{finite}~,
\ee
and
\be\label{Z=}
Z=1+\frac{e_A^2+e_B^2}{8\pi^2\lambda}~\frac{1}{\epsilon}\left(\frac{Zk}{\tilde M}\right)^\epsilon~+~\mbox{finite}~,
\ee
where 
\be\label{dM}
\delta M\equiv \tilde M-M~,
\ee
and dimensional regularisation is used. The latter introduces the arbitrary (transmutation) mass scale $k$ in dimensions $d=4-\epsilon$. 
It is interesting to note that 
the expression for $Z$ does not involve $\omega$, although it does correspond to a non-trivial resumation of higher-order loops.

The SD equation for the dual photon self energy gives the same integral as in the perturbative case, except for a factor $1/Z$
\be\label{o=}
\omega=\frac{e_B^2}{6\pi^2Z}~\frac{1}{\epsilon}\left(\frac{Zk}{\tilde M}\right)^\epsilon~+~\mbox{finite~as}~\epsilon \, \to \, 0^+~,
\ee
which is a non-perturbative effect.

\subsection{Strong coupling limit}

The SD equations relevant to our model lead us to the quantum corrections given by eqs.~\eqref{M=}, \eqref{Z=}) and \eqref{o=}. In a perturbative context, 
these corrections vanish in the limit where the coupling constants go to 0, but we are interested here in a non-trivial solution of the SD equations, which does not
necessarily reproduce the perturbative results for small couplings. In order to achieve this, we first derive the coupled differential equations obtained from
eqs.~\eqref{M=},\eqref{Z=},\eqref{o=}), which describe the evolution of $\tilde M,Z,\omega$ with the scale $k$. Instead of taking boundary conditions set by the perturbative regime,
we explore here the possibility of having $Z=0$ as part of the boundary conditions. We stress that it is only in the framework of effective theories that such a
non-perturbative regime is relevant, and the renormalisation flows we derive in this section are not related to the usual perturbative ones.

Before proceeding, we remark that the usual perturbative renormalisation procedure, which consists of appropriately absorbing the $1/\epsilon$ divergence 
in \eqref{Z=}, introduces the rescaling $\chi_B\to\sqrt{Z_B}\chi_B$,
where $\chi_B$ is the bare fermion field. One chooses then $Z_B\, Z=1$ such that {\it perturbatively},
\be\label{ZB=}
Z_B=1-\frac{e_A^2+e_B^2}{8\pi^2\lambda}~\frac{1}{\epsilon}\left(\frac{Zk}{\tilde M}\right)^\epsilon~+~\cdots~,
\ee
where the dots  represent higher orders in $e_A,e_B$. 
As expected from a consistent unitary description of asymptotic states \cite{psbook}, given the interpretation of the wave-function renormalization as the probability for finding 
such states, the unitarity bound $0\le Z_B<1$ is indeed satisfied. In principle this implies 
\be\label{unitarity}
 Z > 1~, 
\ee
whereas we are interested in the regime where $Z<<1$. The latter regime violates the unitarity bound \eqref{unitarity}, which is possible in the case of {\it composite} monopoles.
As mentioned in the introduction, these are the only type of known solutions of field theories of 
phenomenological interest so far~\cite{thooft,dokos,emy,aruna}. Thus, the effective field theory developed here can {\it only} be associated with {\it composite monopoles/dyons}, 
not necessarily restricted to the aforementioned types, but also encompassing new composite monopole/dyon solutions that may exist in beyond the standard model physics. 

Equation \eqref{Z=} implies
\be
\partial_k Z=\frac{e_A^2+e_B^2}{8\pi^2\lambda}\frac{\tilde M}{Zk}\partial_k\left(\frac{Zk}{\tilde M}\right)~,
\ee
with solution
\be
Z=C_1+\frac{e_A^2+e_B^2}{8\pi^2\lambda}\ln\left(\frac{Zk}{\tilde M}\right)~,
\ee
where $C_1$ is a constant. 
We then introduce the ``running mass'' $M_r(k)$ via the definition 
\be
M_r(k)\equiv\frac{\tilde M}{Z}~,
\ee
from which the scale $k_0$, at which $Z\to0$, can be defined as the solution of the self consistent 
equation\footnote{ The reader should bear in mind that the scale $k_0$ is gauge independent, which can be guaranteed by requiring an appropriate dependence of the constant $C_1$ and 
the bare coupling $e_B$ on the gauge parameter $\lambda$.}
\be\label{self}
k_0=M_0\exp\left(-\frac{8\pi^2\lambda C_1}{e_A^2+e_B^2}\right)~~~~\mbox{where}~~~~M_0\equiv M_r(k_0)~.
\ee

This allows to express $Z$ as
\be\label{Zexpr}
Z=\frac{e_A^2+e_B^2}{8\pi^2\lambda}\ln\left(\frac{kM_0}{k_0M_r(k)}\right)~.
\ee
For the polarisation tensor, we obtain from eq.(\ref{o=}) 
\be
\partial_k(Z\omega)=\frac{e_B^2}{6\pi^2}~\frac{\tilde M}{Zk}\partial_k\left(\frac{Zk}{\tilde M}\right)
~~~~\Rightarrow~~~~Z\omega=C_2+\frac{e_B^2}{6\pi^2}\ln\left(\frac{k}{M_r(k)}\right)~,
\ee
where $C_2$ is a constant. On requiring a finite limit $\omega\to\omega_0$ when $Z\to0$, we can easily determine 
$C_2$, implying, on account of \eqref{Zexpr}, that $\omega$ is actually 
independent of $k$ :
\be\label{omega0}
\omega=\frac{e_B^2}{6\pi^2Z}\ln\left(\frac{kM_0}{k_0M_r(k)}\right)=\frac{4\lambda e_B^2}{3(e_A^2+e_B^2)}=\omega_0~.
\ee 
Notice that $\omega_0\ge0$, since the gauge parameter $\lambda$ is positive.

For the mass correction, we obtain from (\ref{M=})
\be
\partial_k\left(\frac{\delta M}{M_r}\right)
=\frac{\kappa^2}{8\pi^2\lambda}~\frac{M_r}{k}\partial_k\left(\frac{k}{M_r}\right)~,
\ee
where the effective coupling $\kappa$ is given by 
\be
\kappa^2\equiv e_A^2+e_B^2+3\lambda\left(e_A^2+\frac{e_B^2}{1+\omega_0}\right)~.
\ee
As a consequence,
\be
\frac{\delta M}{M_r}=C_3+\frac{\kappa^2}{8\pi^2\lambda}\ln\left(\frac{k}{M_r}\right)~,
\ee
where $C_3$ is a constant. Taking into account the definition $\delta M=\tilde M-M$ and the expression (\ref{Zexpr}) for $Z$, we obtain the self consistent equation 
which must be satisfied by the function $M_r(k)$:
\be\label{equaMr}
\frac{M}{M_r(k)}=\frac{e_A^2+e_B^2-\kappa^2}{8\pi^2\lambda}\ln\left(\frac{k}{M_r(k)}\right)+\frac{e_A^2+e_B^2}{8\pi^2\lambda}\ln\left(\frac{M_0}{k_0}\right)-C_3
\ee
One can check that there is indeed an appropriate solution for $M_0$: the latter equation for $k=k_0$ is of the form
\be
X=-a\ln(X)+b~,
\ee
where $X=k_0/M_0$, $a$ is a positive constant and $b$ is a constant of integration. It can be easily seen graphically that this equation has always a finite and positive solution, 
which ensures the consistency of the present derivation. 

We note that $M_r$ increases with $k$ since, from eq.(\ref{equaMr}),
\be\label{dMdk}
k\partial_k M_r=\frac{(\kappa^2-e_A^2-e_B^2)M_r}{\kappa^2-e_A^2-e_B^2+8\pi^2\lambda M/M_r}
=\frac{3[e_A^2+e_B^2/(1+\omega_0)]M_r}{3[e_A^2+e_B^2/(1+\omega_0)]+8\pi^2 M/M_r}~>0~.
\ee
Also, one can see that for $M > 0$ (and $\omega_0 > 0$)
\be
k\partial_k M_r<M_r~,
\ee
implying that $Z$ increases with $k$:
\be
k\partial_k Z=\frac{e_A^2+e_B^2}{8\pi^2\lambda}\left(1-\frac{k\partial_k M_r}{M_r}\right)>0~.
\ee

The above analysis has shown that it was possible to find a consistent solution of the SD equations, which allowed for the non-perturbative limit $Z\to0$ to be taken. 
Such a solution is parametrised by two scales with dimensions of mass, $k_0$ and $M_0$.
We also note that as $k \to k_0$, the renormalised mass $\tilde M (k) \to \tilde M(k_0)=0$, since $Z\to0$ and $M_r=\tilde M/Z$ goes to the finite value $M_0$. As a consequence, the corrections to the mass of the monopole/dyon are negative when $k\to k_0$, since ({\it cf.} \eqref{dM}) $\delta M(k_0)=-M < 0$. This feature is specific to the non-perturbative regime we consider here, and we emphasise again that it does {\it not}
reduce to the usual perturbative solution when the coupling constant becomes small: it represents instead a {\it disconnected} configuration of quantum fluctuations in the model.

\subsection{Gauge dependence}

The polarisation $\omega_0$ given in the expression (\ref{omega0}) depends on $\lambda$, which is allowed by the non-perturbative resummation provided by the SD equations. 
Unfortunately this has an effect on the gauge-dependence of physical observables. Indeed, let us consider the scalar potential $V$ seen by a dual charge $e_B$, 
and generated by a point-like dual charge at rest, assumed to correspond to a magnetic current $j_\chi^\nu(r)=(e_B\delta(\vec r),0)$, which in Fourier components reads $j_\chi^\nu(q)=(e_B,0)$. 
We have 
\be\label{vg}
V= e_B \, B_0 = e_B\, \Delta_{0\nu}^B j_\chi^\nu~,
\ee
where $\Delta_{\mu\nu}^B$ is the dual photon propagator \eqref{photonprops}. For the static modes $q^\mu=(0,\vec q)$, this leads to the Coulomb potential 
\be
V(\vec r)= e_B \, \int\frac{d^3q}{(2\pi)^3}\, \Delta^B_{00}(q) \, j^0_\chi(q) \, e^{i\vec q\cdot\vec r} =\frac{e_B^2}{1+\omega_0}\int\frac{d^3q}{(2\pi)^3}\frac{e^{i\vec q\cdot\vec r}}{(\vec q)^2}
=\frac{e_B^2}{4\pi |\vec r|(1+\omega_0)}~,
\ee
which depends on the gauge through $\omega_0$.

A similar situation characterises dynamically generated masses in non perturbative gauge theories~\cite{am}. 
Nevertheless, it is possible that, through appropriate resummation of special classes of appropriate Feynman graphs (``pinched technique''~\cite{pt}),  
one recovers the gauge-independent value of physical quantities, such as masses, or in our case dressed charges, 
and such values seem to correspond to the value one would obtain in the truncated SD treatment in Feynman gauge $\lambda=1$. In what follows, we therefore make this gauge choice, 
and we identify the effective dual coupling
\be\label{eBtilde}
\tilde e_B^2\equiv\frac{e_B^2}{1+\omega_0(\lambda=1)} =\frac{3\,e_B^2\, (e_A^2+e_B^2)}{3\,e_A^2+7\,e_B^2}~,
\ee
from which the flow for the running mass given in eq.(\ref{dMdk}) reads 
\be\label{gaugeindepM}
k\partial_k M_r=\frac{(e_A^2+\tilde e_B^2)M_r}{e_A^2+\tilde e_B^2+(8\pi^2/3)M/M_r}~,
\ee
an will be used in the discussion below.

\section{Electron-monopole scattering and an effective velocity-dependent magnetic charge for (composite) monopoles}\label{emscat}

As we have discussed in the introduction, one may obtain the differential cross section for non-relativistic electron-monopole scattering at small angles using  
quantum mechanical treatment, leading to the expression \eqref{diffcs}, which scales with the relative momentum $p= \mu v$ of the incident matter particle  
(in the frame where the heavy monopole is initially at rest)  as $1/p^2$ instead of the traditional $1/p^4$ of the conventional Ruhterford scattering differential 
cross section \eqref{Rutherford1} between charged matter particles. As already mentioned previously, upon invoking electric-magnetic duality, one may obtain the Rutherford 
scattering formula from \eqref{diffcs}, by replacing the monopole magnetic coupling with an effective velocity($\beta$)-dependent magnetic coupling, \eqref{geff1}, \eqref{defbeta}.
In what follows, we propose an origin of such a coupling based on the non-perturbative results derived in the previous section.

\begin{figure}[h]
 \centering
 \includegraphics[clip,width=0.40\textwidth]{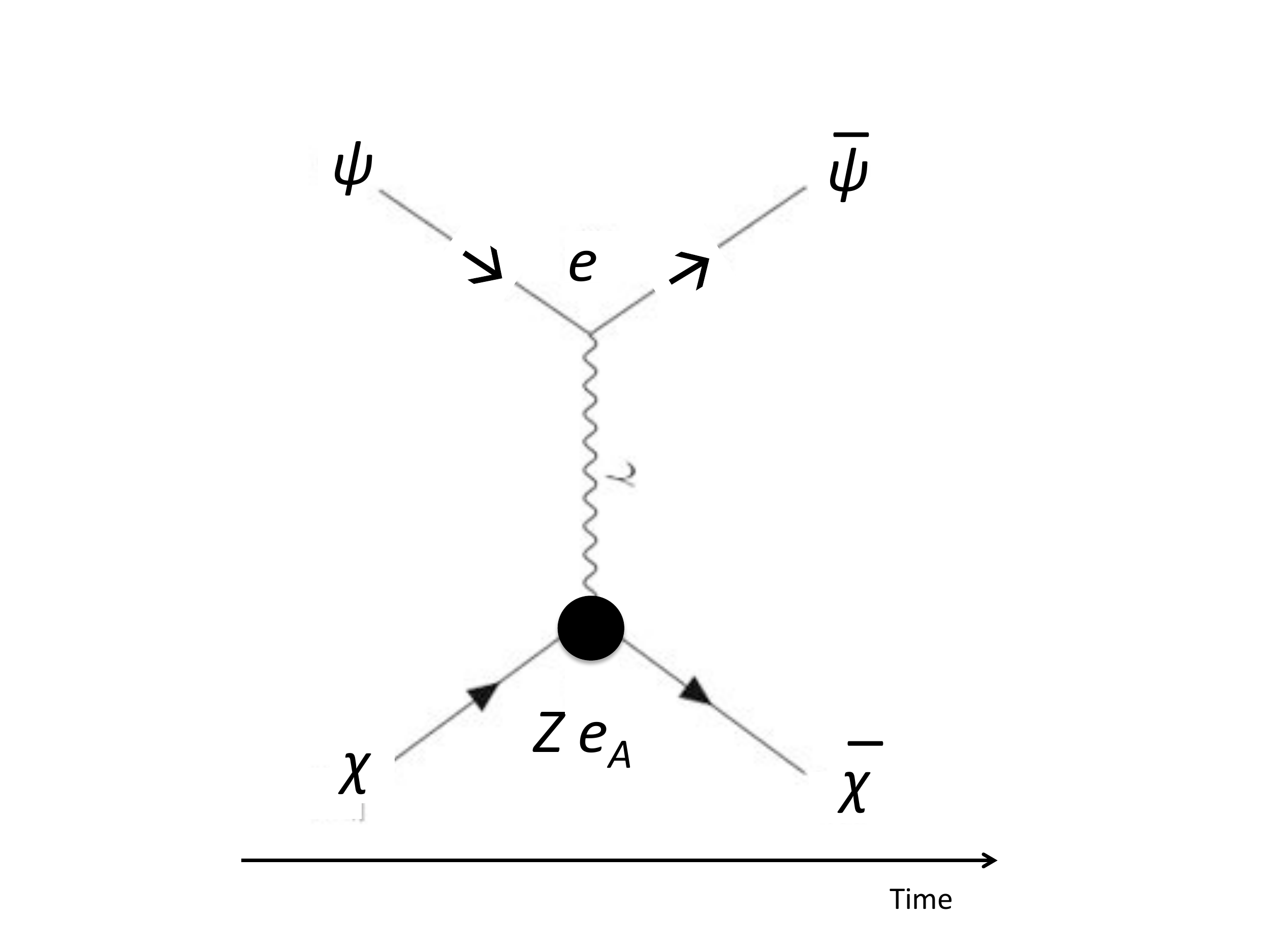} 
\caption{Typical scattering graph between an electron ($\psi$) and a fermion-monopole ($\chi$). Wavy lines denote electrodynamical photons, $\gamma$. Straight lines with arrows denote the 
scattered fermions (dashed arrow lines denote electrons, continuous arrow lines denote the monopole. The dark blob denotes the dressed coupling $Ze_A$ due to quantum corrections induced by the strongly
coupled dual photon, which only the monopole couples to, characterised by the presence of the $Z$ factor. By rotating the graph by 90$^o$ counterclockwise, one obtains a Drell-Yan diagram 
for monopole-antimonopole pair production from matter fermions.}
\label{fig:scat}
\end{figure}

\subsection{Perturbative Effective Magnetic charge for slowly-moving (composite) monopoles}

To this end, we consider {\it slowly moving} monopoles, relative to the incident matter particle, assumed for concreteness to be an electron of charge $e$. 
The one-photon exchange scattering Feynman graph of fig.~\ref{fig:scat}, between the fermions $\psi$ (electron) and $\chi$ (monopole fermion), will prove sufficient for this purpose, 
as a result of the perturbative nature of both  couplings, the electron charge $e \ll 1$, but also the coupling of the monopole $\chi$ to the real photon. 
The latter is determined by the appropriate dressed vertex 
$\Lambda_\mu^{A\chi}=e_A Z \, \gamma_\mu$, \eqref{vertices}, with the wave function $Z$ to be determined from \eqref{Zexpr}, and as we shall see below  it can become perturbative, 
when the scale $k$ approaches  $k_0$, for which $Z \ll 1$, with $Z(k_0) = 0$. We shall also be able to express the magnetic charge $g$ of the monopole, 
which is subjected to the quantization condition \eqref{schquant}, in terms of $e$.

We first notice that, from the differential equation (\ref{gaugeindepM}), one can expand $M_r$ to first order in $k-k_0$  to find
\be
\frac{M_r(k)}{M}\simeq\frac{M_0}{M}+a\frac{k-k_0}{k_0}~,
\ee
where
\be
a\equiv\frac{(e_A^2+\tilde e_B^2)(M_0/M)^2}{8\pi^2/3+(e_A^2+\tilde e_B^2)~M_0/M}
\ee
From the expression (\ref{Zexpr}) for $Z$ we obtain then, to first order in $k-k_0$,
\be\label{Zkk0}
Z\simeq \frac{Z_0}{k_0}(k-k_0)~,
\ee
where
\be\label{Zfinal}
Z_0=\frac{e_A^2+e_B^2}{8\pi^2+3(e_A^2+\tilde e_B^2)M_0/M}~,
\ee
with $\tilde e_B$ given by \eqref{eBtilde}. 

The Feynman diagram of fig.~\ref{fig:scat} makes sense if the dressed coupling of the monopole to the electromagnetic photon $\gamma$ is small, 
which necessitates $Z < 1$  in \eqref{Zkk0}, and can be achieved for $k \to k_0$.  
Then the scattering process resembles that of two ordinary 
charged particles in (perturbative) electrodynamics with charges $e$ and $Z\, e_A$, both small. 
The result for the corresponding differential cross section can 
be borrowed from standard quantum electrodynamics then. Since our interest is to compare this result with the differential cross section \eqref{diffcs} of 
the non relativistic scattering of an electron off a 
monopole with magnetic charge $g$, discussed in \cite{milton},  we consider here the appropriate limit, of a low relative velocity 
between the electron and the monopole. The result from the differential cross section at small angles $\theta $ , in a frame where the monopole is initially at rest, 
reads then (in appropriate units in which the fine structure constant $\alpha = e^2$, as per the conventions of \cite{milton})~\cite{qedbook}:
\begin{equation}\label{ddscat}
\frac{d \sigma}{d \Omega}|_{\rm LAB} \simeq \left(\frac{ Z \, e_A\, e }{2 \mu |\vec{v}|^2 {\rm sin}^2 \frac{\theta}{2}}\right)^2 \, \stackrel{\theta \ll 1}{\simeq}  \,
\left(\frac{ Z_0 \frac{k - k_0}{k_0} \, e_A\, e }{2 \mu |\vec{v}|^2 }\right)^2 \, \frac{1}{(\theta/2)^4}~.
\end{equation}
where $\mu$ is the reduced mass of the electron-monopole system. 

Comparing \eqref{ddscat} with \eqref{diffcs}, which as we discussed in section \ref{sec:intro} describes quantum scattering of an electron off a dyon at small angles, 
we then observe that one can define 
an effective magnetic charge, describing the coupling of the monopole to the photon,  by (\emph{cf.} \eqref{Zkk0})
\be\label{geff4}
g_{eff}=Ze_A \simeq Z_0\, \frac{k-k_0}{k_0} \, e_A~, 
\ee
with the scale $|k-k_0|$ being identified with a proper (Lorentz invariant) centre-of-mass momentum scale, and $k_0$ is identified with a mass scale through the self-consistent relation (\ref{self}),
where we choose the constant of integration $C_1$ such that
\be
k_0=2M_0~.
\ee
Consistently with the effective field theory approach for monopole-antimonopole pair production~\cite{monprod}, we may take the mass scale $M_0$ to represent the monopole (rest) mass, and in this way one  can reproduce the formula \eqref{diffcs} of \cite{milton},
and arrive at an effective ``velocity-dependent'' magnetic charge (\ref{geff1})\footnote{The reader should bear in mind the Lorentz-invariant nature of the scale \eqref{cor}, 
characteristic of the effective field theory~\cite{monprod}, which was not evident in the initial non-relativistic quantum mechanical approach to dyon-dyon scattering of \cite{milton}.}
\begin{equation}\label{cor}
\frac{|k - k_0|}{k_0} \rightarrow \frac{\sqrt{E^2 - 4M^2_0}}{2M_0} = \frac{E}{2M_0}\, \sqrt{1 - \frac{4M_0^2}{s}} \simeq \beta,
\end{equation}
with $E^2=s=(p_1+p_2)^2$ the relevant Mandelstam variable (with $p_1$ ($p_2$) the incoming (outgoing) four momenta in the scattering process of fig.~\ref{fig:scat} 
(or the corresponding fused momenta in the case of production). In such an identification, $E$  denotes the exchanged photon momentum scale (see fig.~\eqref{fig:scat}). 
In the last approximate equality of \eqref{cor}, we considered 
slowly moving monopoles, for which $E \simeq 2M_0$ (since their kinetic energy is small), and this case corresponds to $ k \to k_0$ 
where perturbativity applies, and 
\begin{equation}\label{geff2}
g_{eff} \simeq Z_0 \, \beta \,  e_A~,
\end{equation}
with $Z_0$ given by \eqref{Zfinal}. In the duality-compatible approach of \cite{milton}, $g_{eff}=g \beta$, with $g$ the magnetic charge that in the monopole case satisfies 
the appropriate quantization rule \eqref{schquant}. Hence, it is natural to identify in the context of our effective field theory
\begin{equation}\label{magneticcharge}
g=Z_0 \, e_A~.
\end{equation}
Schwinger's quantization condition for the Monopole \eqref{schquant} would then require 
\begin{equation}\label{zquant}
Z_0 \, e_A =  \frac{n}{\alpha} \, e~, \quad  n \in {\tt Z}~,
\end{equation} 
with ${\tt Z}$ the set of integers, and $\alpha$ the fine structure constant of electromagnetism. 

The value of $Z_0$ depends on several parameters, and the quantization condition \eqref{zquant} is not compatible with all of them. 
For bare couplings $e_B^2 \gg e_A^2$, we note that we also have $\tilde e_B^2 \simeq 3e_B^2/7\gg e_A^2$
and we can see, for instance, that the case $M_0\gg M$, would be inconsistent with \eqref{zquant}. On the other hand, 
for $M_0 \ll M$, we have  from \eqref{Zfinal},
\begin{equation}\label{Z2}
Z_0 \simeq \frac{7\tilde e_B^2}{24\, \pi^2},  ~~~~~(M_0 \ll M)~,
\end{equation}
which, together with eq.~\eqref{zquant}, implies a quantization condition for the (gauge independent) dressed dual photon coupling 
$\tilde e_B$ \eqref{eBtilde}. In this sense, we understand why the dual photon coupling is proportional to the magnetic charge of the monopole, 
and how our effective theory \eqref{lag} describes the scattering of the latter with charged matter or its production from charged matter~\cite{milton}. 

At this point we make some important remarks, which relate to the fact that the above effective field theory approach can {\it only} describe {\it composite} monopoles, 
like the ones that exist currently as topological soliton solutions of several field theories of phenomenological relevance~\cite{thooft,dokos,emy,aruna}, as discussed in 
the introduction of the article. This becomes clear by noticing that, because of the presence of the transmutation scale $k$, the wave function renormalization $Z$ \eqref{Zkk0} 
for the monopole fermion can, and in fact it does for $k  \to k_0$, become smaller than unity.  This violates the unitarity bound~\cite{psbook} that requires  the wave function
renormalization to be larger than one ({\it cf.} \eqref{unitarity}), should the monopole be an elementary particle asymptotic state. However, such a bound can be evaded for {\it composite} states, either of 
the type discussed in the literature so far, or new, yet unknown structured solutions to be discovered in theories beyond the standard model. 

We stress that in our approach, the arguments of \cite{drukier} on the strong suppression of the cross section for the collider production of such (composite) monopole-antimonopole 
pairs by form factors 
of order ${\mathcal O}\Big(e^{-\frac{4}{\alpha}}\Big) \simeq 10^{-238}$, are not valid for slowly moving monopoles; effectively, such factors would be 
replaced by terms of order ${\mathcal O}\Big(e^{-\frac{4}{\alpha}\, \frac{k-k_0}{k_0}}\Big)$ in our case, which would be of ${\mathcal O}(1)$ for slowly moving heavy monopoles ($k \to k_0$), with 
masses much larger than the SM quarks or leptons, whose collisions can lead to the monopole production.  
At present, such remarks of course remain speculations, as we do not have a complete microscopic understanding on the strong $U(1)$ dual interactions involved in our effective model.

\begin{figure}[t]
 \centering
 \includegraphics[clip,width=0.70\textwidth]{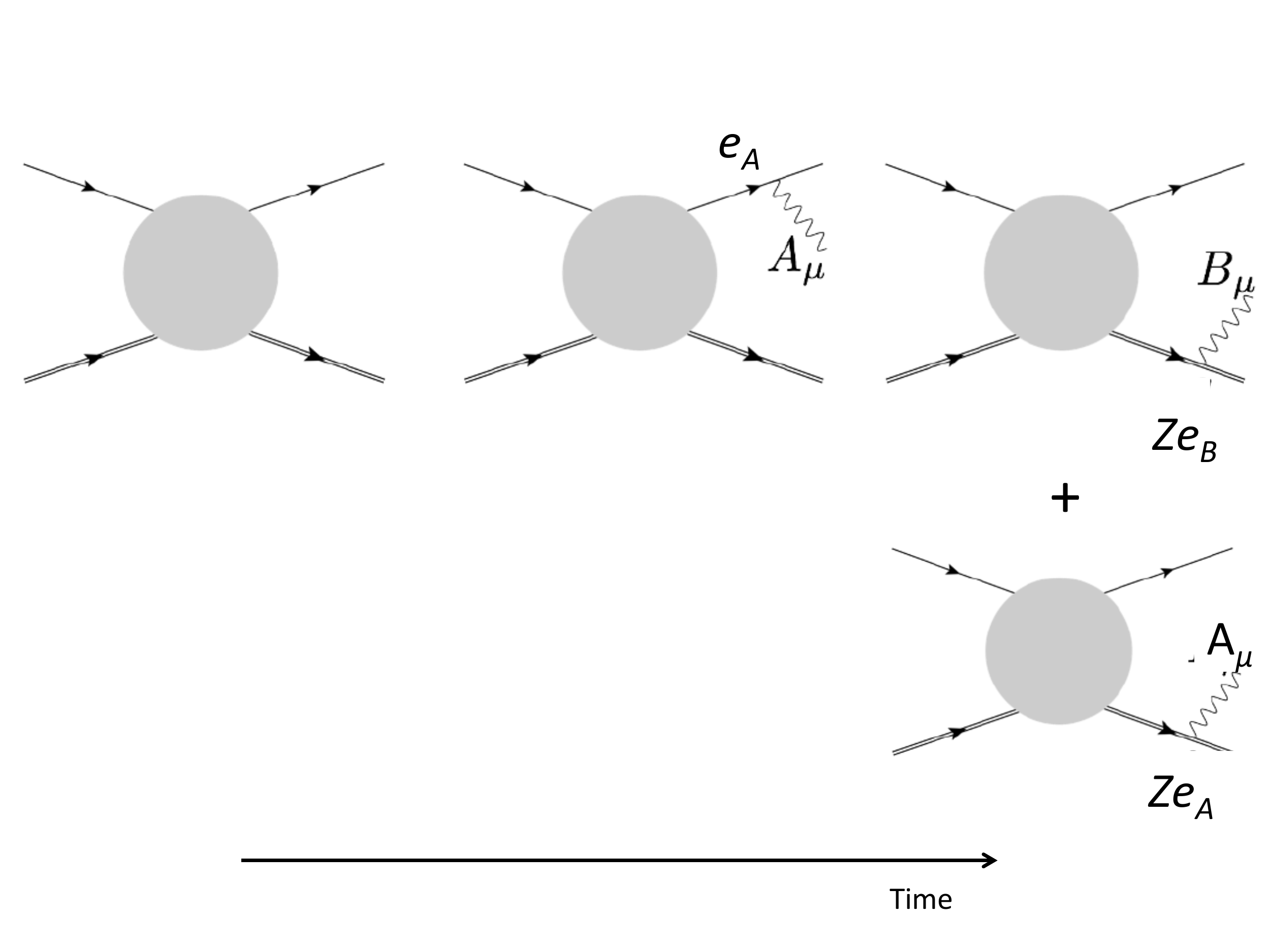} 
 \caption{Typical scattering graphs between an electron (representing generically charged matter) and a fermion-monopole. Wavy lines denote real (on-shell) 
soft electrodynamical photons $A_\mu$, or gauge bosons 
$B^\mu$ of the strongly-coupled 
U(1)$_{\rm strong}$, the latter associated only with the monopole/dyon. Thin straight lines with arrows denote the 
scattered electrons, whilst double arrowed straight lines denote the monopole/dyon. The gray blobs denote generic scattering processes, involving all fields. 
The dressed couplings of the monopole to the $A^\mu$ (photon) and  $B^\mu$ (dual-photon) fields  are given by $Ze_A$ and $Ze_B$ respectively, 
where $Z$ is the wave-function renormalization factor, computed non-perturbatively due to the quantum corrections induced by the strongly-coupled $U(1)_{\rm strong}$ gauge interactions. 
Notice that in our approach, in contrast to that of Zwanziger~\cite{Zwanziger:1970hk}, which was adopted in \cite{terning}, 
the photon ($A^\mu$) and the dual photon ($B^\mu$) are independent fields, which explains the presence of the last two graphs on the right-hand side of the figure (connected with a ``+'' sign). 
When however consider the monopole, one should impose on the classical on-shell gauge fields the constraint \eqref{solconstr}, which brings back the Dirac-string effects. 
For slowly moving monopoles, $Z \ll 1$ and this ensures perturbativity of all couplings, hence the depicted graphs denote the leading corrections in both $A^\mu$ and $B^\mu$ sectors.
In such a case, resummation of the soft on-shell photons exponentiates such string effects into phases of the pertinent scattering amplitudes~\cite{terning}.}
\label{fig:IR}
\end{figure}

\subsection{Soft gauge-field-emission resummation and disappearance of Dirac-String effects from cross sections: a brief review of the arguments of \cite{terning}}

Before closing the discussion, we would like to review briefly the arguments of \cite{terning} on the decoupling of any Lorentz-violating Dirac-string effects from the physical cross sections
(or even the scattering amplitudes if the quantization condition \eqref{schquant} is valid). These arguments were based on a perturbative resummation of soft photons in the toy model for 
the monopole used in that work, where the magnetic charge was perturbative. This was achieved in \cite{terning} by assuming dark sector monopoles, and weak coupling of the monopole sector 
with the visible sector. In our case, as discussed above, both the magnetic and electric charge couplings of the monopole to photon and dual photon are {\it perturbative} for 
{\it slowly moving} monopoles, due to the wave-function renormalization screening effects on the effective magnetic charge \eqref{vertices}, \eqref{geff2}, 
(see also  \eqref{magneticcharge}-\eqref{Z2}).

For such perturbative couplings, the leading soft gauge field (photons and dual photons) emissions that affect generic scattering processes of electrons off fermion-monopoles in our effective $U(1)_{\rm weak} \times U(1)_{\rm strong}$ gauge theory are depicted in figure \ref{fig:IR}. The analysis has been performed in \cite{terning} and in previous relevant literature referenced there, and we shall not repeat it here. We only point out a trivial modification in our case as compared to those works, namely the addition of the soft-$B^\mu$(dual-photon)- emission graph, since, in contrast to the  standard Zwanziger approach~\cite{Zwanziger:1970hk}, here the fields $A^\mu$ and $B^\mu$ are treated as independent gauge fields, not related through \eqref{relAB}, 
except when one considers classical solutions \eqref{solconstr}. Such solutions are the on-shell solutions represented by the external gauge field wavy lines in the graphs, and will combine the $A$ and $B$ photon emission into one; this is indicated with the pertinent ``+'' sign in  fig.~\eqref{fig:IR}.

The presence of a classical solution, satisfying the constraint \eqref{solconstr} implies a classical Lagrangian \eqref{lag} which has Chern-Simons-like topological terms {\it mixing}
the ${\mathcal A}$ and ${\mathcal B}$ fields, which depend on the Lorentz-Violating Dirac-string effects ($\propto \eta^\mu$ fixed four-vector). 
However, upon resumming soft photon and dual photon emissions,  such effects exponentiate in the pertinent scattering amplitude, 
in such a way that they appear {\it only} in the phases of the amplitude, as demonstrated in \cite{terning}, 
whose arguments apply here as a consequence of our perturbative magnetic charge couplings. Upon the quantization condition \eqref{schquant}, 
which should be respected by the non perturbative dressing, as discussed above, the effects disappear also from the phase of the scattering amplitude. 
This completes our discussion on the self consistency of our perturbative effective gauge field theory approach to the (slow) monopole-matter scattering or 
(slow)monopole-antimonopole-pair-production-from-matter processes, which have been conjectured in the previous literature and used in monopole searches at colliders~\cite{moedal,monprod}

\section{Conclusions and Outlook}\label{concl}

In this work we have considered a rather toy model of magnetically-charged spin-$\frac{1}{2}$ ``dyons" interacting with
ordinary electrically-charged matter fermions (electrons for concreteness). We have considered an effective gauge field theory $U(1)_{\rm weak} \times U(1)_{\rm strong}$, 
where the weak $U(1)$ represents 
electromagnetism, whilst the strong $U(1)$ effectively describes interactions that would lead to the presence of magnetic poles in then theory. 
The presence of two potentials was adopted by Zwanziger~\cite{Zwanziger:1970hk} in an attempt to describe the effective theory of Dirac monopoles in a local fashion, 
avoiding the use of infinite strings. However, the theory lacked Lorentz invariance. In a recent work~\cite{terning}, employing a toy model of perturbative magnetic charge, 
it was argued that all such string effects can be resummed to a physically-irrelevant phase of the scattering amplitude between monopoles and matter, 
and in case the charge quantization condition applies, such phases vanish, so the amplitude itself was Lorentz (and gauge) invariant. 
In realistic monopole models, perturbative magnetic charges have been argued by Schwinger, Milton and collaborators~\cite{milton} 
to  characterise  the interaction of 
monopoles with matter as a result of electric magnetic duality. However, 
such arguments were purely qualitative, not supported by any detailed modeling. 
They were based on quantum mechanical scattering of electrons off magnetic poles, whose cross sections had a form that, 
in order to be obtained from Rutherford cross section of normal charged particles by means of imposing electromagnetic duality, 
one should impose that the magnetic charge was effectively proportional to the monopole velocity, hence perturbative for slowly moving monopoles.

Motivated by these results, we have discussed in this work a toy model for describing the interaction of magnetic monopoles/dyons with matter, 
where quantum fluctuation effects of the magnetic charge are represented by the presence of a second independent strong U(1) gauge potential. However, 
we have ignored any Lorentz (or gauge) violation effects,
anticipating the results of \cite{terning}. Our analysis has lead to a consistent result, namely, the emergence of an effective magnetic coupling of the monopole/dyon to the real photons, 
depending on the monopole velocity, in the spirit conjectured in \cite{milton}. This result was obtained as a mathematically-self-consistent solution of the appropriate Schwinger Dyson equations. 
The role of the velocity-dependent factor that appeared in the work of \cite{milton} is played here by the transmutation mass scale that 
appears in the Schwinger-Dyson solution for the monopole wave-function renormalization, which enters the expression of the monopole-photon vertex. 
The role of the non-perturbative corrections of the dual U(1)$_{\rm strong}$ is crucial to this effect. However,  we stress that in the case of slowly-moving monopoles/dyons,
the wave-function renormalization violates unitarity bounds set for elementary particle asymptotic states \eqref{unitarity}, thus necessitating the application of our effective 
U(1)$_{\rm weak} \times {\rm U(1)}_{\rm strong}$ gauge field theory {\it only} to {\it composite} monopoles/dyons, of generic type though, not necessarily restricted to the known 
solutions existing in the literature.

Once such a perturbative monopole-photon coupling is established (for slowly moving, heavy (compared to the matter-fermion mass) monopoles), 
then the arguments of \cite{terning} are in operation, and one may assume that any Lorentz non-invariant string effects that might appear in the theory of 
\cite{Zwanziger:1970hk} affect {\it only} the phases of the respective scattering amplitudes, and they actually disappear once the charge quantization condition \eqref{schquant} is imposed.  
In our approach we have explained how such a quantization arises naturally for the dual photon coupling. 
Although in the theory of \cite{Zwanziger:1970hk} such string effects implied that the dual photon is not independent of the real photon, 
and that on-shell there are only two propagating degrees of freedom, in our fully quantum theory, the dual photon quantum fluctuations have been assumed independent of electromagnetism. 
Nonetheless, one may impose the monopole solution of Zwanziger as a consistent Lorentz non invariant solution to the respective equations of motion, 
thus making contact with the original Dirac monopole case. 

We should stress though that the emergence of the  strong U(1) interactions from microscopic considerations in concrete composite monopole models existing in the current literature~\cite{thooft,dokos,emy,aruna} is still not understood. It might be possible that the very existence of a magnetic pole implies automatically such interactions, as an effective way of describing the monopole quantum fluctuations.\footnote{In this latter respect, we mention that an interesting research direction would be to attempt and find connections, if any, of the current approach with the manifestly dual quantum field theory proposal for electric and magnetic charges of \cite{marchetti}, in which the Dirac string (and thus the associated fixed vector $\eta^\mu$ in the Zwanziger approach~\cite{Zwanziger:1970hk}) becomes dynamical.}
Nonetheless, our effective field theory description of monopole-matter interactions developed in the current 
work offers support to the relativistic effective gauge field theory approach of \cite{monprod}, used so far in collider searches of magnetically charged particles~\cite{moedal}. Formulating our theory on the lattice might then be a way forward for obtaining results in the non-perturbative regime of the effective magnetic couplings, when the monopoles are not slowly moving, or appear as virtual particles in quantum loops of, say, light-by-light scattering processes, of interest to collider searches for monopoles heavier than the respective collider production threshold. In  the latter case, an enhanced light-by-light scattering cross section, as compared to the Standard-Model case, has been claimed in ref.~\cite{ginzburg}, as an indirect way of detecting the existence of monopoles. In that work, analytical estimates for the corresponding cross sections have been made on assuming {\it ad hoc} photon-energy dependent monopole-photon couplings of the form \eqref{geff4},\eqref{cor}, which are weak for sufficiently low photon energies compared to the monopole mass. Our effective gauge field theory not only provides justification for such assumptions, as we discussed above, but, by  being placed on the lattice, can also lead, in principle, to an extension of the considerations of \cite{ginzburg} to higher photon energies and lower monopole masses, thus enhancing the appropriate parameter space region in future collider searches. 
Such issues constitute interesting research material for future work.

\section*{Acknowledgements}

It is a pleasure to acknowledge discussions with fellow members of the MoEDAL Collaboration. 
The work of JA and NEM is supported in part by the UK Science and Technology Facilities  research Council (STFC) under the research grant
ST/P000258/1. NEM also acknowledges a scientific associateship (``\emph{Doctor Vinculado}'') at IFIC-CSIC-Valencia University, Valencia, Spain.

\appendix

\section{One-Particle-Irreducible (1PI) effective action \label{appA}}

\numberwithin{equation}{section}

\setcounter{equation}{0}

\nin$\bullet$ The partition function is
\be\label{partition}
\mathcal Z=\int{\cal D}[\phi]\exp\left(iI+iI_s\right)~,
\ee
where $\phi$ denotes collectively the different degrees of freedom $A_\mu,B_\mu,\psi\,\ovl\psi,\chi,\ovl\chi$ with sources $j_A^\mu,j_B^\mu,\ovl\eta,\eta,\ovl\xi,\xi$ respectively, such that
the source term is
\be
I_s=\int_x j_A^\mu A_\mu+j_B^\mu B_\mu+\ovl\eta\psi+\ovl\psi\eta+\ovl\xi\chi+\ovl\chi\xi~.
\ee
The connected-graphs generating functional is
\be
W=-i\ln \mathcal Z~,
\ee
whose functional derivatives define the background fields
\bea\label{background}
&&\psi_b=\left<\psi\right>=\frac{\delta W}{\delta\ovl\eta}~~~~,~~~~\ovl\psi_b=\left<\ovl\psi\right>=\frac{\delta W}{\delta\eta}\\
&&\chi_b=\left<\chi\right>=\frac{\delta W}{\delta\ovl\xi}~~~~,~~~~\ovl\chi_b=\left<\ovl\chi\right>=\frac{\delta W}{\delta\xi}\nn
&&A^\mu_b=\left<A^\mu\right>=\frac{\delta W}{\delta j_\mu^A}~~~~,~~~~B^\mu_b=\left<B^\mu\right>=\frac{\delta W}{\delta j_\mu^B}~,\nonumber
\eea
with
\be
\left<\cdots\right>\equiv\frac{1}{\mathcal Z}\int{\cal D}[\phi]\left(\cdots\right)\exp\left(iI+iI_s\right)~.
\ee

\nin$\bullet$ The one-particle-irreducible(1PI)-graphs generating functional, the effective action $\Gamma$, is a functional of the background fields 
and is obtained via a Legendre transform of $W$,
after inverting the relations (\ref{background}) and treating the sources as functionals of the background fields
\be
\Gamma=W-\int_x j_A^\mu A_\mu^b+j_B^\mu B_\mu^b+\ovl\eta\psi_b+\ovl\psi_b\eta+\ovl\xi\chi_b+\ovl\chi_b\xi~.
\ee
$\Gamma$ has the following functional derivatives
\bea\label{1stGdif}
&&\frac{\delta\Gamma}{\delta\psi_b}=-\ovl\eta~~~~,~~~~\frac{\delta\Gamma}{\delta\ovl\psi_b}=-\eta\\
&&\frac{\delta\Gamma}{\delta\chi_b}=-\ovl\xi~~~~,~~~~\frac{\delta\Gamma}{\delta\ovl\chi_b}=-\xi\nn
&&\frac{\delta\Gamma}{\delta A_b^\mu}=-j_\mu^A~~~~,~~~~\frac{\delta\Gamma}{\delta B_b^\mu}=-j_\mu^B~,\nonumber
\eea

\nin$\bullet$ The bare propagators are defined by
\bea
&&S_\psi=\left(\frac{\delta^2 I}{\delta\psi\delta\ovl\psi}\right)_0^{-1}~~~~,~~~~S_\chi=\left(\frac{\delta^2 I}{\delta\chi\delta\ovl\chi}\right)_0^{-1}\\
&&D_{\mu\nu}=\left(\frac{\delta^2 I}{\delta A^\nu\delta A^\mu}\right)_0^{-1}=\left(\frac{\delta^2 I}{\delta B^\nu\delta B^\mu}\right)_0^{-1}~,\nonumber
\eea
where the subscript 0 denotes vanishing fields. As explained in section \ref{Furry}, the dressed propagators are diagonal, both for fermions and gauge fields, and are defined by
\bea
&&G_\psi=\left(\frac{\delta^2\Gamma}{\delta\psi_b\delta\ovl\psi_b}\right)_0^{-1}~~~~,~~~~G_\chi=\left(\frac{\delta^2\Gamma}{\delta\chi_b\delta\ovl\chi_b}\right)_0^{-1}\\
&&\Delta_{\mu\nu}^A=\left(\frac{\delta^2 \Gamma}{\delta A_b^\nu\delta A_b^\mu}\right)_0^{-1}~~~~,~~~~\Delta_{\mu\nu}^B=\left(\frac{\delta^2 \Gamma}{\delta B_b^\nu\delta B_b^\mu}\right)_0^{-1}~.\nonumber
\eea 
The dressed vertices are
\bea
&&\Lambda_\mu^{A\psi}=\left.\frac{\delta^3\Gamma}{\delta\psi_b\delta A^\mu_b\delta\ovl\psi_b}\right|_0~~~~,~~~~
\Lambda_\mu^{A\chi}=\left.\frac{\delta^3\Gamma}{\delta\chi_b\delta A^\mu_b\delta\ovl\chi_b}\right|_0\\
&&\Lambda_\mu^{B\psi}=\left.\frac{\delta^3\Gamma}{\delta\psi_b\delta B^\mu_b\delta\ovl\psi_b}\right|_0~~~~,~~~~
\Lambda_\mu^{B\chi}=\left.\frac{\delta^3\Gamma}{\delta\chi_b\delta B^\mu_b\delta\ovl\chi_b}\right|_0~.\nonumber
\eea

\nin$\bullet$ From the first functional derivatives (\ref{1stGdif}), one can obtain the following identities, relating the second functional derivatives of $W$ and $\Gamma$
\bea
&&\frac{\delta^2\Gamma}{\delta\psi_b\delta\ovl\psi_b}=-(\delta^2W)^{-1}_{\eta,\ovl\eta}~~~~,~~~~
\frac{\delta^2\Gamma}{\delta\chi_b\delta\ovl\chi_b}=-(\delta^2W)^{-1}_{\xi\ovl\xi}\\
&&\frac{\delta^2\Gamma}{\delta A_b^\mu\delta A_b^\nu}=-(\delta^2W)^{-1}_{j_\mu^A j_\nu^A}~~~~,~~~~
\frac{\delta^2\Gamma}{\delta B_b^\mu\delta B_b^\nu}=-(\delta^2W)^{-1}_{j_\mu^B j_\nu^B}~,\nonumber
\eea
where the inverse has to be read as the element of the inverse matrix.

\section{Ward identities \label{appB}}

\numberwithin{equation}{section}

\setcounter{equation}{0}

\nin$\bullet$ The gauge transformation (\ref{gauge}) leaves the action $S$ invariant and modifies the source terms, such that the partition function (\ref{partition})
is modified as
\bea
\mathcal Z&\to& \mathcal Z+i\int{\cal D}[\phi]\exp\left(iI+iI_s\right)\\
&&\times\int_x\Big(j_A^\mu \partial_\mu\theta_A+j_B^\mu \partial_\mu\theta_B
+ie\theta_A(\ovl\eta\psi-\ovl\psi\eta)+i(e_A\theta_A+e_B\theta_B)(\ovl\xi\chi-\ovl\chi\xi)\Big)\nn
&&~~~~~~~~~~~~~~~~~~~~~~~~~~~~~~~~~~~~~~~~~+{\cal O}(\theta^{A,B})^2~.\nonumber
\eea
Integration by parts lead to
\bea
\mathcal Z&\to& \mathcal Z+i\int_x\theta_A\int{\cal D}[\phi]\exp\left(iI+iI_s\right)\Big(-\partial_\mu j_A^\mu+ie(\ovl\eta\psi-\ovl\psi\eta)+ie_A(\ovl\xi\chi-\ovl\chi\xi)\Big)\\
&&~~~~+i\int_x\theta_B\int{\cal D}[\phi]\exp\left(iI+iI_s\right)\Big(-\partial_\mu j_B^\mu+ie_B(\ovl\xi\chi-\ovl\chi\xi)\Big)\nn
&&~~~~~~+{\cal O}(\theta^{A,B})^2~,\nonumber
\eea
and gauge invariance for independent functions $\theta_A$ and $\theta_B$ leads to
\bea
0&=&\int{\cal D}[\phi]\exp\left(iI+iI_s\right)\Big(-\partial_\mu j_A^\mu+ie(\ovl\eta\psi-\ovl\psi\eta)+ie_A(\ovl\xi\chi-\ovl\chi\xi)\Big)\\
0&=&\int{\cal D}[\phi]\exp\left(iI+iI_s\right)\Big(-\partial_\mu j_B^\mu+ie_B(\ovl\xi\chi-\ovl\chi\xi)\Big)\nonumber~.
\eea
Together with the identities given in Appendix \ref{appA}, we obtain then
\bea\label{basics}
0&=&\partial_\mu \left(\frac{\delta\Gamma}{\delta A_\mu}\right)+ie\frac{\delta\Gamma}{\delta\psi_b}\psi_b-ie\ovl\psi_b\frac{\delta\Gamma}{\delta\ovl\psi_b}
+ie_A\frac{\delta\Gamma}{\delta\chi_b}\chi_b-ie_A\ovl\chi_b\frac{\delta\Gamma}{\delta\ovl\chi_b}\\
0&=&\partial_\mu \left(\frac{\delta\Gamma}{\delta B_\mu}\right)+ie_B\frac{\delta\Gamma}{\delta\chi_b}\chi-ie_B\ovl\chi\frac{\delta\Gamma}{\delta\ovl\chi_b}~.\nonumber
\eea

\nin $\bullet$ On account of the property (\ref{nomixpsichi}), taking the functional derivatives of eqs.(\ref{basics}) with respect to $\psi_b$ and $\ovl\psi_b$, 
followed by setting the background fields to 0, leads to
\bea
\frac{1}{e}\frac{\partial\Lambda^{A\psi}_\mu}{\partial z_\mu} (x,y,z)&=&iG_\psi^{-1}(x,z)\delta(x-y)-iG_\psi^{-1}(y,z)\delta(x-y)\\
\frac{\partial\Lambda^{B\psi}_\mu}{\partial z_\mu}(x,y,z)&=&0~.\nonumber
\eea

\nin$\bullet$ For the same reason, on taking the functional derivatives of eqs.(\ref{basics}) with respect to $\chi_b$ and $\ovl\chi_b$, and then setting the background fields to 0, leads to
\bea
\frac{1}{e_A}\frac{\partial\Lambda^{A\chi}_\mu}{\partial z_\mu} (x,y,z)&=&iG_\chi^{-1}(x,z)\delta(x-y)-iG_\chi^{-1}(y,z)\delta(x-y)\\
\frac{1}{e_B}\frac{\partial\Lambda^{B\chi}_\mu}{\partial z_\mu} (x,y,z)&=&iG_\chi^{-1}(x,z)\delta(x-y)-iG_\chi^{-1}(y,z)\delta(x-y)~.\nonumber
\eea
In Fourier components, we have
\be
iG^{-1}(x,z)-iG^{-1}(y,z)~~~\longrightarrow~~~G^{-1}(p+q)-G^{-1}(p)=q^\mu\left.\frac{\partial G^{-1}}{\partial p^\mu}\right|_p+{\cal O}(q^2)~.
\ee
where $p$ is an incoming fermion momentum and $q$ is the gauge field momentum. In the limit $q\to0$, we then obtain the Ward identities
\be
\frac{1}{e}\Lambda^{A\psi}_\mu(p,0)=\frac{\partial G_\psi^{-1}}{\partial p^\mu}~~~~,~~~~\frac{1}{e_B}\Lambda^{B\chi}_\mu(p,0)=\frac{\partial G_\chi^{-1}}{\partial p^\mu}~,
\ee
and the additional properties
\be\label{additionalB}
q^\mu \Lambda^{B\psi}_\mu(p,q)=0~~~~,~~~~e_B q^\mu\Lambda^{A\chi}_\mu(p,q)=e_A q^\mu\Lambda^{B\chi}_\mu(p,q)~.
\ee

\section{Schwinger-Dyson (SD) equations \label{appC}}

\numberwithin{equation}{section}

\setcounter{equation}{0}

\nin$\bullet$ The SD equation for the $\psi$-self energy is obtained by noting that the integral of the following functional derivative vanishes
\be
\int{\cal D}[\phi]\frac{\delta }{\delta\ovl\psi}\exp(iI+iI_s)=0~,
\ee
such that
\be
\int{\cal D}[\phi]\left(i\gamma^\mu D^A_\mu\psi-m\psi+\eta\right)\exp(iI+iI_s)=0~.
\ee
Using the identities in Appendix \ref{appA}, the latter equation can also be written
\be
0=\left(i\gamma^\mu\partial_\mu-m\right)\psi_b-\frac{\delta\Gamma}{\delta\ovl\psi_b}-e\gamma^\mu\left<A_\mu\psi\right>~,
\ee
and a functional derivative with respect to $\psi_b$, followed by setting the fields to 0, leads to
\be
G_\psi^{-1}-S_\psi^{-1}=-e\gamma^\mu\frac{\delta}{\delta\psi_b}\left<A_\mu\psi\right>_0~.
\ee
We have 
\be
\left<A_\mu\psi\right>=A_\mu^b\psi_b-i\frac{\delta^2W}{\delta j^\mu_A\delta\ovl\eta}~,
\ee
and
\be
\left.\frac{\delta}{\delta\psi_b}\frac{\delta^2W}{\delta j^\mu_A\delta\ovl\eta}\right|_0
=-\left.\frac{\delta^3W}{\delta\eta\delta j^\mu_A\delta\ovl\eta}\right|_0G_\psi^{-1}~,
\ee
where the missing terms vanish when fields are set to 0, and the integral over spacetime coordinates is understood. The next step consists in noting that
\bea
\left.\frac{\delta^3W}{\delta\eta\delta j^\mu_A\delta\ovl\eta}\right|_0&=&-\left.\frac{\delta}{\delta j^\mu_A}\left(\frac{\delta^2\Gamma}{\delta\psi_b\delta\ovl\psi_b}\right)^{-1}\right|_0\\
&=&G_\psi\left(\frac{\delta^3\Gamma}{\delta\psi_b\delta A_\nu\delta\ovl\psi_b}\frac{\delta A_\nu}{\delta j^\mu_A}+
\frac{\delta^3\Gamma}{\delta\psi_b\delta B_\nu\delta\ovl\psi_b}\frac{\delta B_\nu}{\delta j^\mu_A}\right)_0G_\psi\nn
&=&G_\psi\Lambda_{A\psi}^\nu\Delta_{\mu\nu}^AG_\psi~,
\eea
where $\delta B_\nu/\delta j^A_\mu|_0=0$ since the gauge propagator is diagonal. We finally obtain
\be
G_\psi^{-1}-S_\psi^{-1}=ie\gamma^\mu G_\psi\Lambda_{A\psi}^\nu\Delta_{\mu\nu}^A~,
\ee
where the integration over space time coordinates in understood.\\

\nin$\bullet$ The SD equation for the $\chi$-fermion self-energy is obtained by noting that 
\be
\int{\cal D}[\phi]\frac{\delta }{\delta\ovl\chi}\exp(iI+iI_s)=0~,
\ee
such that
\be
\int{\cal D}[\phi]\left(i\gamma^\mu D^{A+B}_\mu\chi-M\chi+\xi\right)\exp(iI+iI_s)=0~,
\ee
and the same steps as the ones described above lead to 
\be
G_\chi^{-1}-S_\chi^{-1}=ie_A\gamma^\mu G_\chi\Lambda_{A\chi}^\nu\Delta_{\mu\nu}^A+ie_B\gamma^\mu G_\chi\Lambda_{B\chi}^\nu\Delta_{\mu\nu}^B~.
\ee
\\

\nin $\bullet$ The SD equation for the $A_\mu$-self energy is obtained by noting that 
\be
\int{\cal D}[\phi]\frac{\delta }{\delta A_\mu}\exp(iI+iI_s)=0~,
\ee
such that
\be\label{equaDA}
D_{\mu\nu}^{-1} A^\nu_b-e\left<\ovl\psi\gamma_\mu\psi\right>-e_A\left<\ovl\chi\gamma_\mu\chi\right>+j_\mu^A=0~.
\ee
A functional derivative with respect to $A^\nu_b$ leads to
\be
\left(\Delta^A_{\mu\nu}\right)^{-1}-D_{\mu\nu}^{-1}=ie~\mbox{tr}\left\{\gamma_\mu\frac{\delta}{\delta A^\nu_b}\frac{\delta^2W}{\delta\eta\delta\ovl\eta}\right\}
+ie_A~\mbox{tr}\left\{\gamma_\mu\frac{\delta}{\delta A^\nu_b}\frac{\delta^2W}{\delta\xi\delta\ovl\xi}\right\}
\ee
and for vanishing fields we have
\bea
&&\left.\frac{\delta}{\delta A^\nu_b}\frac{\delta^2W}{\delta\eta\delta\ovl\eta}\right|_0
=\frac{\delta}{\delta A^\nu_b}\left(\frac{\delta^2\Gamma}{\delta\psi_b\delta\ovl\psi_b}\right)_0^{-1}
=-G_\psi\Lambda^{A\psi}_\nu G_\psi\\
&&\left.\frac{\delta}{\delta A^\nu_b}\frac{\delta^2W}{\delta\xi\delta\ovl\xi}\right|_0
=\frac{\delta}{\delta A^\nu_b}\left(\frac{\delta^2\Gamma}{\delta\psi_b\delta\ovl\psi_b}\right)_0^{-1}
=-G_\chi\Lambda^{A\chi}_\nu G_\chi~.
\eea
Finally we obtain
\be
\left(\Delta^A_{\mu\nu}\right)^{-1}-D_{\mu\nu}^{-1}=i~\mbox{tr}\left\{e~\gamma_\mu G_\psi\Lambda^{A\psi}_\nu G_\psi+e_A~\gamma_\mu G_\chi\Lambda^{A\chi}_\nu G_\chi\right\}~.
\ee
\\

\nin$\bullet$ The SD equation for the $B_\mu$-self energy is obtained by noting that 
\be
\int{\cal D}[\phi]\frac{\delta }{\delta B_\mu}\exp(iI+iI_s)=0~,
\ee
such that
\be\label{equaDB}
D_{\mu\nu}^{-1} B^\nu_b-e_B\left<\ovl\chi\gamma_\mu\chi\right>+j_\mu^B=0~,
\ee
and similar steps as the ones described above lead to
\be
\left(\Delta^{-1}_{\mu\nu}\right)^B-D_{\mu\nu}^{-1}=ie_B~\mbox{tr}\left\{\gamma_\mu G_\chi\Lambda^{B\chi}_\nu G_\chi\right\}~.
\ee
\\

\section{Calculation of quantum corrections \label{appD}}

\numberwithin{equation}{section}

\setcounter{equation}{0}

\nin$\bullet$ Within the approximations described in Section \ref{nonperturb}, the SD equation for the monopole self energy is
\bea\label{SDM}
(Z-1)\slashed p-\delta M&=&ie_A^2\int\frac{d^4q}{(2\pi)^4}\gamma^\mu G_\chi(p+q)Z\gamma^\nu D_{\mu\nu}(q)\\
&&+ie_B^2\int\frac{d^4q}{(2\pi)^4}\gamma^\mu G_\chi(p+q)Z\gamma^\nu\Delta_{\mu\nu}^B(q)\nonumber
\eea
and the SD equation for the dual photon polarisation tensor is
\be\label{SDB}
\omega\left(q^2\eta_{\mu\nu}-q_\mu q_\nu\right)=ie_B^2~\mbox{tr}\int\frac{d^4p}{(2\pi)^4}\gamma_\mu G_\chi(p)Z\gamma_\nu G_\chi(p+q)~,
\ee
where the propagators and vertices are given in section \ref{nonperturb}. \\

\nin$\bullet$ For the monopole self energy (\ref{SDM}), the mass corrections is obtained by setting the external momentum to 0
\bea
\delta M&=&ie_A^2k^\epsilon\int\frac{d^dq}{(2\pi)^d}\gamma^\mu\frac{\tilde M}{Z^2q^2-\tilde M^2}
\gamma^\nu\left(\frac{\eta_{\mu\nu}}{q^2}+\frac{1-\lambda}{\lambda}\frac{q_\mu q_\nu}{q^4}\right)\\
&&+ie_B^2k^\epsilon\int\frac{d^dq}{(2\pi)^d}\gamma^\mu\frac{\tilde M}{Z^2q^2-\tilde M^2}
\frac{Z\gamma^\nu}{1+\omega}\left(\frac{\eta_{\mu\nu}}{q^2}+\frac{1+\omega-\lambda}{\lambda}\frac{q_\mu q_\nu}{q^4}\right)~.\nonumber
\eea
and the wave function renormalisation is obtained by taking a derivative with respect to the external momentum component $p^\sigma$, which is then set to 0
\bea\label{Zindep}
&&(Z-1)\gamma_\sigma\\
&=&ie_A^2k^\epsilon\int\frac{d^dq}{(2\pi)^d}\gamma^\mu\left(\frac{Z\gamma_\sigma}{Z^2q^2-\tilde M^2}-\frac{2Z^2(Z\slashed q-\tilde M)q_\sigma}{(Z^2q^2-\tilde M^2)^2}\right)
\gamma^\nu\left(\frac{\eta_{\mu\nu}}{q^2}+\frac{1-\lambda}{\lambda}\frac{q_\mu q_\nu}{q^4}\right)\nonumber\\
&&+ie_B^2k^\epsilon\int\frac{d^dq}{(2\pi)^d}\gamma^\mu\left(\frac{Z\gamma_\sigma}{Z^2q^2-\tilde M^2}-\frac{2Z^2(Z\slashed q-\tilde M)q_\sigma}{(Z^2q^2-\tilde M^2)^2}\right)
\frac{Z\gamma^\nu}{1+\omega}\left(\frac{\eta_{\mu\nu}}{q^2}+\frac{1+\omega-\lambda}{\lambda}\frac{q_\mu q_\nu}{q^4}\right)~.\nonumber
\eea
In the latter expressions dimensional regularisation is used, in dimension $d=4-\epsilon$ and with the arbitrary mass scale $k$. 
In both cases the simplification of the integrand is straightforward, and we obtain for the mass corrections 
\bea
\frac{\delta M}{\tilde M}&=&\frac{i}{\lambda}\left(e_A^2(1+3\lambda)+e_B^2\frac{1+3\lambda+\omega}{1+\omega}\right)
~\frac{k^\epsilon}{Z^{d-3}}\int\frac{d^dq}{(2\pi)^d}\frac{1}{q^2(q^2-\tilde M^2)}~+~\mbox{finite}\\
&=&\frac{1}{8\pi^2\lambda Z}\left(e_A^2(1+3\lambda)+e_B^2\frac{1+3\lambda+\omega}{1+\omega}\right)~\frac{1}{\epsilon}\left(\frac{Zk}{\tilde M}\right)^\epsilon~+~\mbox{finite}~,\nonumber
\eea

For the monopole wave function renormalisation, we observe that both $Z$ and $\omega$ cancel in the integrand (both in the diverging and finite parts), 
such that the corrections arising from $A_\mu$ are identical to those arising from $B_\mu$, and we obtain 
\bea
Z&=&1-i\frac{e_A^2+e_B^2}{\lambda}~\frac{k^\epsilon}{Z^{d-4}}\int\frac{d^dq}{(2\pi)^d}\frac{1}{(q^2-\tilde M^2)^2}~+~\mbox{finite}\\
&=&1+\frac{e_A^2+e_B^2}{8\pi^2\lambda}~\frac{1}{\epsilon}\left(\frac{Zk}{\tilde M}\right)^\epsilon~+~\mbox{finite}~,\nonumber
\eea

\nin$\bullet$ For the dual photon polarisation, the integral in eq.(\ref{SDB}) is the same as in the usual perturbative case, except for an overall factor $1/Z$, 
which is a non-perturbative feature:
\be
\omega=\frac{e_B^2}{6\pi^2Z}~\frac{1}{\epsilon}\left(\frac{Zk}{\tilde M}\right)^\epsilon~+~\mbox{finite}~.
\ee


\begin{thebibliography}{99}

\bibitem{Dirac}
  P.~A.~M.~Dirac,
  Phys.\ Rev.\  {\bf 74}, 817 (1948).
  doi:10.1103/PhysRev.74.817



\bibitem{Zwanziger:1970hk} 
  D.~Zwanziger,
  Phys.\ Rev.\ D {\bf 3}, 880 (1971).
  doi:10.1103/PhysRevD.3.880

\bibitem{Weinberg}
  S.~Weinberg,
  Phys.\ Rev.\  {\bf 138} (1965) B988.
  doi:10.1103/PhysRev.138.B988;


\bibitem{schw} 
  J.~S.~Schwinger,
  Phys.\ Rev.\  {\bf 144}, 1087 (1966).
  doi:10.1103/PhysRev.144.1087
  Phys.\ Rev.\  {\bf 173}, 1536 (1968).
  doi:10.1103/PhysRev.173.1536;
  Phys.\ Rev.\ D {\bf 12}, 3105 (1975).
  doi:10.1103/PhysRevD.12.3105

\bibitem{thooft}
  G.~'t Hooft,
  Nucl.\ Phys.\ B {\bf 79} (1974) 276.
  doi:10.1016/0550-3213(74)90486-6
A.~M.~Polyakov,
  JETP Lett.\  {\bf 20}, 194 (1974)
  [Pisma Zh.\ Eksp.\ Teor.\ Fiz.\  {\bf 20}, 430 (1974)].



\bibitem{gg}
  H.~Georgi and S.~L.~Glashow,
  Phys.\ Rev.\ Lett.\  {\bf 32} (1974) 438.
  doi:10.1103/PhysRevLett.32.438
  
  
\bibitem{dokos} 
  C.~P.~Dokos and T.~N.~Tomaras,
  Phys.\ Rev.\ D {\bf 21}, 2940 (1980).
  doi:10.1103/PhysRevD.21.2940
  
\bibitem{patrizii} 
  L.~Patrizii and M.~Spurio,
  ``Status of Searches for Magnetic Monopoles,''
  Ann.\ Rev.\ Nucl.\ Part.\ Sci.\  {\bf 65}, 279 (2015)
  doi:10.1146/annurev-nucl-102014-022137
  [arXiv:1510.07125 [hep-ex]], and references therein;
 V.~A.~Mitsou,
  ``The quest for magnetic monopoles ? past, present and future,''
  PoS CORFU {\bf 2017}, 188 (2018).
  doi:10.22323/1.318.0188, and references therein.
   
  
\bibitem{wen} 
  X.~G.~Wen and E.~Witten,
  Nucl.\ Phys.\ B {\bf 261}, 651 (1985).
  doi:10.1016/0550-3213(85)90592-9
  
  
\bibitem{shafi} 
  T.~W.~Kephart, G.~K.~Leontaris and Q.~Shafi,
  JHEP {\bf 1710}, 176 (2017)
  doi:10.1007/JHEP10(2017)176
  [arXiv:1707.08067 [hep-ph]].

  

 
 \bibitem{cho}  Y.~M.~Cho and D.~Maison,
  Phys.\ Lett.\ B {\bf 391}, 360 (1997)
  doi:10.1016/S0370-2693(96)01492-X
  [hep-th/9601028].
 
 
 \bibitem{emy} Y.~M.~Cho, K.~Kim and J.~H.~Yoon,
  Eur.\ Phys.\ J.\ C {\bf 75}, no. 2, 67 (2015)
  doi:10.1140/epjc/s10052-015-3290-3
  [arXiv:1305.1699 [hep-ph]];
 J.~Ellis, N.~E.~Mavromatos and T.~You,
  Phys.\ Lett.\ B {\bf 756}, 29 (2016)
  doi:10.1016/j.physletb.2016.02.048
  [arXiv:1602.01745 [hep-ph]].
 
 \bibitem{aruna} S.~Arunasalam and A.~Kobakhidze,
  Eur.\ Phys.\ J.\ C {\bf 77}, no. 7, 444 (2017)
  doi:10.1140/epjc/s10052-017-4999-y
  [arXiv:1702.04068 [hep-ph]];
 J.~Ellis, N.~E.~Mavromatos and T.~You,
  Phys.\ Rev.\ Lett.\  {\bf 118}, no. 26, 261802 (2017)
  doi:10.1103/PhysRevLett.118.261802
  [arXiv:1703.08450 [hep-ph]].
  
\bibitem{sarkar} N.~E.~Mavromatos and S.~Sarkar,
  Phys.\ Rev.\ D {\bf 95}, no. 10, 104025 (2017)
  doi:10.1103/PhysRevD.95.104025
  [arXiv:1607.01315 [hep-th]];
  Phys.\ Rev.\ D {\bf 97}, no. 12, 125010 (2018)
  doi:10.1103/PhysRevD.97.125010
  [arXiv:1804.01702 [hep-th]].
  Universe {\bf 5}, no. 1, 8 (2018)
  doi:10.3390/universe5010008
  [arXiv:1812.00495 [hep-ph]].

  
  \bibitem{monprod} 
Y.~Kurochkin, I.~Satsunkevich, D.~Shoukavy, N.~Rusakovich and Y.~Kulchitsky,
  Mod.\ Phys.\ Lett.\ A {\bf 21}, 2873 (2006).
  doi:10.1142/S0217732306022237;
T.~Dougall and S.~D.~Wick,
  Eur.\ Phys.\ J.\ A {\bf 39} (2009) 213
  doi:10.1140/epja/i2008-10701-8
  [arXiv:0706.1042 [hep-ph]].
   L.~N.~Epele, H.~Fanchiotti, C.~A.~G.~Canal, V.~A.~Mitsou and V.~Vento,
  Eur.\ Phys.\ J.\ Plus {\bf 127}, 60 (2012)
  doi:10.1140/epjp/i2012-12060-8
  [arXiv:1205.6120 [hep-ph]];
S.~Baines, N.~E.~Mavromatos, V.~A.~Mitsou, J.~L.~Pinfold and A.~Santra,
  Eur.\ Phys.\ J.\ C {\bf 78}, no. 11, 966 (2018)
  Erratum: [Eur.\ Phys.\ J.\ C {\bf 79}, no. 2, 166 (2019)]
  doi:10.1140/epjc/s10052-018-6440-6, 10.1140/epjc/s10052-019-6678-7
  [arXiv:1808.08942 [hep-ph]].
  
\bibitem{drukier}
  A.~K.~Drukier and S.~Nussinov,
  Phys.\ Rev.\ Lett.\  {\bf 49} (1982) 102.
  doi:10.1103/PhysRevLett.49.102

\bibitem{psbook} C.~Itzykson and J.~B.~Zuber,
  ``Quantum Field Theory,''
  New York, Usa: Mcgraw-hill (1980) (International Series In Pure and Applied Physics);
 M.~E.~Peskin and D.~V.~Schroeder,
  ``An Introduction to quantum field theory,''
  (Reading, USA: Addison-Wesley (1995))
    ISBN: 9780201503975, 0201503972.



\bibitem{terning} 
  J.~Terning and C.~B.~Verhaaren,
  JHEP {\bf 1903}, 177 (2019)
  doi:10.1007/JHEP03(2019)177
  [arXiv:1809.05102 [hep-th]];
  JHEP {\bf 1812}, 123 (2018)
  doi:10.1007/JHEP12(2018)123
  [arXiv:1808.09459 [hep-th]].





\bibitem{milton} J.~S.~Schwinger, K.~A.~Milton, W.~y.~Tsai, L.~L.~DeRaad, Jr. and D.~C.~Clark,
  Annals Phys.\  {\bf 101}, 451 (1976).
  doi:10.1016/0003-4916(76)90020-8;
for a review see: 
  K.~A.~Milton,
  Rept.\ Prog.\ Phys.\  {\bf 69}, 1637 (2006)
  doi:10.1088/0034-4885/69/6/R02
  [hep-ex/0602040], and references therein.



\bibitem{moedal}
  A.~Abulencia {\it et al.} [CDF Collaboration],
  Phys.\ Rev.\ Lett.\  {\bf 96}, 201801 (2006)
  doi:10.1103/PhysRevLett.96.201801
  [hep-ex/0509015];
G.~Aad {\it et al.} [ATLAS Collaboration],
  Phys.\ Rev.\ Lett.\  {\bf 109}, 261803 (2012)
  doi:10.1103/PhysRevLett.109.261803
  [arXiv:1207.6411 [hep-ex]];
  arXiv:1905.10130 [hep-ex];
  B.~Acharya {\it et al.} [MoEDAL Collaboration],
  Phys.\ Lett.\ B {\bf 782}, 510 (2018)
  doi:10.1016/j.physletb.2018.05.069
  [arXiv:1712.09849 [hep-ex]];
  B.~Acharya {\it et al.} [MoEDAL Collaboration],
  [arXiv:1903.08491 [hep-ex]].
  


\bibitem{cptmag} See, for instance: 
  P.~S.~Bisht, T.~Li, Pushpa and O.~P.~S.~Negi,
  Int.\ J.\ Theor.\ Phys.\  {\bf 49}, 1370 (2010)
  doi:10.1007/s10773-010-0317-2
  [arXiv:0911.2341 [hep-th]].



\bibitem{am}  See, for instance, J.~Alexandre,
  arXiv:1009.5834 [hep-ph];
N.~E.~Mavromatos,
  Phys.\ Rev.\ D {\bf 83}, 025018 (2011)
  doi:10.1103/PhysRevD.83.025018
  [arXiv:1011.3528 [hep-ph]] and references therein.

\bibitem{pt}  J.~M.~Cornwall and J.~Papavassiliou,
  Phys.\ Rev.\ D {\bf 40}, 3474 (1989).
  doi:10.1103/PhysRevD.40.3474
J.~Papavassiliou,
  Phys.\ Rev.\ D {\bf 41}, 3179 (1990).
  doi:10.1103/PhysRevD.41.3179
D.~Binosi and J.~Papavassiliou,
  J.\ Phys.\ G {\bf 30}, 203 (2004)
  doi:10.1088/0954-3899/30/2/017
  [hep-ph/0301096];
  Phys.\ Rept.\  {\bf 479}, 1 (2009)
  doi:10.1016/j.physrep.2009.05.001
  [arXiv:0909.2536 [hep-ph]], and references therein.


\bibitem{qedbook} See, for example,  I.~J.~R.~Aitchison and A.~J.~G.~Hey,
  {\it Gauge theories in particle physics: A practical introduction. Vol. 1: From relativistic quantum mechanics to QED} (CRC Press (2012), 
    Bristol, UK: IOP (2003) ISBN: 9781466512993)

\bibitem{marchetti} 
  K.~Lechner and P.~A.~Marchetti,
  Nucl.\ Phys.\ B {\bf 569}, 529 (2000)
  doi:10.1016/S0550-3213(99)00711-7
  [hep-th/9906079].


\bibitem{ginzburg} 
  I.~F.~Ginzburg and A.~Schiller,
  Phys.\ Rev.\ D {\bf 60}, 075016 (1999)
  doi:10.1103/PhysRevD.60.075016
  [hep-ph/9903314];
  Phys.\ Rev.\ D {\bf 57}, 6599 (1998)
  doi:10.1103/PhysRevD.57.R6599
  [hep-ph/9802310].


\end{thebibliography}
\end{document}